\documentclass[twocolumn]{aa}
\usepackage{graphicx}
\usepackage{natbib}
\graphicspath{{./Fig/}}

\usepackage{bbm}
\usepackage{pstricks}
\usepackage{amsmath}
\usepackage{xspace}

	\def\beqx{$$}						\def\eeqx{$$}
	\def\bcc{\begin{center}}		\def\ecc{\end{center}}
	\def\beqnx{\begin{eqnarray*}}	\def\eeqnx{\end{eqnarray*}}
	\def\beqn{\begin{eqnarray}}	\def\eeqn{\end{eqnarray}}

	\def\Weight{\vec{W}}
	\def\Cvre{\vec{T}}
	\def\Conv{\vec{H}}
	\def\Doux{\vec{D}}
	\def\Id{\vec{I}}
	\def\Fou{\vec{F}}

	\def\Lambdab{\vec{\Lambda}_\mathrm{D}}

	\def\DeltaC{\vec{\Delta}_\mathrm{C}}
	\def\DeltaP{\vec{\Delta}_\mathrm{P}}
	\def\DeltaE{\vec{\Delta}_\mathrm{E}}
	
	\def\MatP{\vec{M}_\mathrm{P}}
	\def\MatE{\vec{M}_\mathrm{E}}

	\def\Data{\vec{y}}
	\def\Bruit{\vec{b}}
	\def\Objb{\vec{x}} \def\Obj{x} 
	\def\ObjbES{\vec{x}_\mathrm{e}}\def\ObjES{x_\mathrm{e}} 
	\def\ObjbPS{\vec{x}_\mathrm{p}}\def\ObjPS{x_\mathrm{p}} 
	
	\def\Dom{{\cal M}}
	\def\Supp{{\cal S}}

	\def\cD{\mathrm{c}}
	\def\sD{\mathrm{s}}
	\def\mD{\mathrm{m}}
	\def\dD{\mathrm{d}}
	\def\eD{\mathrm{e}}
	\def\pD{\mathrm{p}}

	\def\Lagr{\ell}
	\def\Lagrb{\vec{\ell}}
	\def\Slack{s}
	\def\Slackb{\vec{s}}
	\def\LagrFun{{\cal L}}

	\def\LS{^{\mathrm{\scriptscriptstyle LS}}}
	\def\LSBi{^{\mathrm{\scriptscriptstyle LS}}_{\mathrm{\scriptscriptstyle Mix}}}
	\def\RegBi{^{\mathrm{\scriptscriptstyle Reg}}_{\mathrm{\scriptscriptstyle Mix}}}

	\def\eC{\mathbbm{C}}
	\def\eR{\mathbbm{R}}
	\def\Unbb{\mathbbm{1}}
	\def\percent{\%}

	\def\Tr{^{\mathrm{t}}}
	\def\eps{\varepsilon}
	\def\rond#1{\overset{\kern-0.33em~_\circ}{#1}}

	\def\aprio{\textit{a priori}\xspace}
	\def\eg{\textit{e.g.}\xspace}
	\def\ie{\textit{i.e.}\xspace}
	\def\PQSC{{(\cal P)}}
	\def\wrt{w.r.t.\xspace}

	\def\dps{\displaystyle}
	\def\incirc#1{\pscirclebox[framesep=1pt]{\scriptsize#1}}
\begin{document}

\title{Positive deconvolution for superimposed \\extended source and point sources.}
\titlerunning{Positive deconvolution for ES + SP.}

\author{J.-F. Giovannelli\inst{1} \and A. Coulais\inst{2}}


\institute{Groupe Probl\`emes inverses, Laboratoire des Signaux et Syst\`emes (\textsc{cnrs}~--~Sup\'elec~--~\textsc{ups}), Plateau de Moulon, 91192 Gif-sur-Yvette, France, \email{giova@lss.supelec.fr} \and   Laboratoire d'\'Etude du Rayonnement de la Mati\`ere en Astrophysique (\textsc{lerma}), Observatoire de Paris, 61 Avenue de l'Observatoire, 75014 Paris, France, \email{alain.coulais@obspm.fr}}

\date{Submitted January, 2004 and revised January, 2005}

\abstract{
The paper deals with the construction of images from visibilities acquired using aperture synthesis instruments: Fourier synthesis, deconvolution, and spectral interpolation/extrapolation. Its intended application is to specific situations in which the imaged object possesses two superimposed components: ($i$) an extended component together with ($ii$) a set of point sources. It is also specifically designed to the case of positive maps, and accounts for a known support. Its originality lies within joint estimation of the two components, coherently with  data, properties of each component, positivity and possible support. We approach the subject as an inverse problem within a regularization framework: a regularized least-squares criterion is specifically proposed and the estimated maps are defined as its minimizer. We have investigated several options for the numerical minimization and we propose a new efficient algorithm based on augmented Lagrangian. Evaluation is carried out using simulated and real data (from radio interferometry) demonstrating the capability  to accurately separate the two components. 
}
 
\maketitle

\section{Introduction}
\nocite{Daniell80,Buck89,Macaulay89,Macaulay94,Schwartz78,Lannes97}

Radio interferometers can be seen as instruments measuring a set of 2D-Fourier coefficients (visibilities) of the brightness distribution of a region in the sky. Visibilities are measured in the Fourier domain (the $(u,v)$-plane) by means of different baselines (projected distance between cross-correlated antennas). Practically, there are two principal deficiencies~\citep{Thompson01} in the visibilities 

\begin{enumerate}

\item the limited coverage of the  $(u,v)$-plane, 

\item measurements errors (especially in millimeter range). 

\end{enumerate}
Regarding point~1, three limitations are encountered. 
\begin{itemize}

\item Usually the central part of the aperture (up to the antenna diameter) is not observed. From this stand point, interferometers behave as high pass filters.

\item Information above the longest baseline is unavailable. In this sense the instruments behave as low pass filters. 

\item The $(u,v)$-plane coverage is irregular, especially when there is a small number of antennas. This results in dirty beam (Fourier transform of visibility weights) with intricate structure and strong sidelobes.

\end{itemize}
Thus, such instruments can be seen as  band pass filters with an intricate impulse response (dirty beam), and noisy output. As a consequence, the available data is relatively poor for imaging objects with various spatial structures extended over the whole frequency domain. In order to compensate for these deficiencies, a large number of methods (from model fitting to non parametric deconvolution) has been continuously proposed (see review in~\citep{Starck02}) and specialized for different types of maps. The present paper deals with a particular type of map consisting of the superimposition of two components.
\begin{itemize}

\item \emph{Point Source} (PS),  or nearly black objects: essentially null-component, with a few strong point sources. 

\item \emph{Extended Sources} (ES): spatially extended, smooth components. 

\end{itemize}
The problem at hand is to build reliable and accurate estimates of two distinct maps (one for PS, one for ES) from a unique given set of visibilities. The question arises \eg for radio imaging of the solar corona at meter wavelength where very strong storms are superimposed over a more stable and large quiet Sun radio-emission (see Sect.~\ref{Sec:NRH}).

\begin{figure}[htb]
\bcc
\begin{tabular}{cc}
\rotateleft{~~~~~~~~\rotateright{\textit{(a)}}}	& \includegraphics[width=7.5cm]{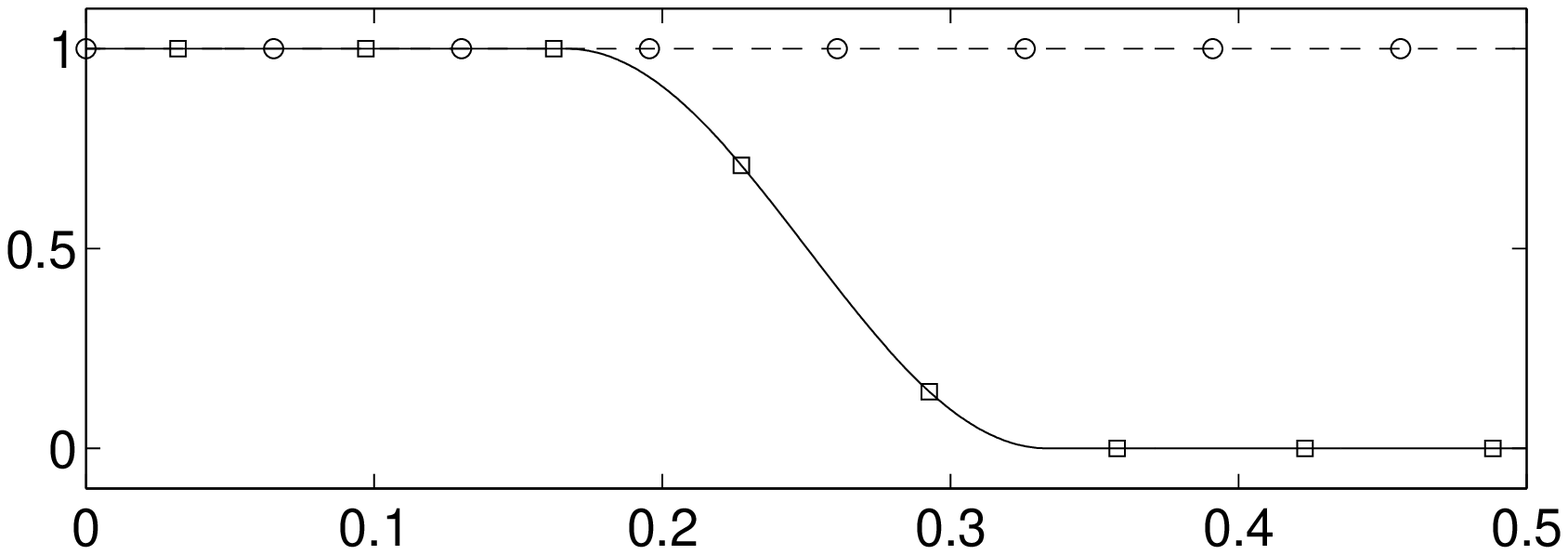}\\
\rotateleft{~~~~~~~~\rotateright{\textit{(b)}}}	& \includegraphics[width=7.5cm]{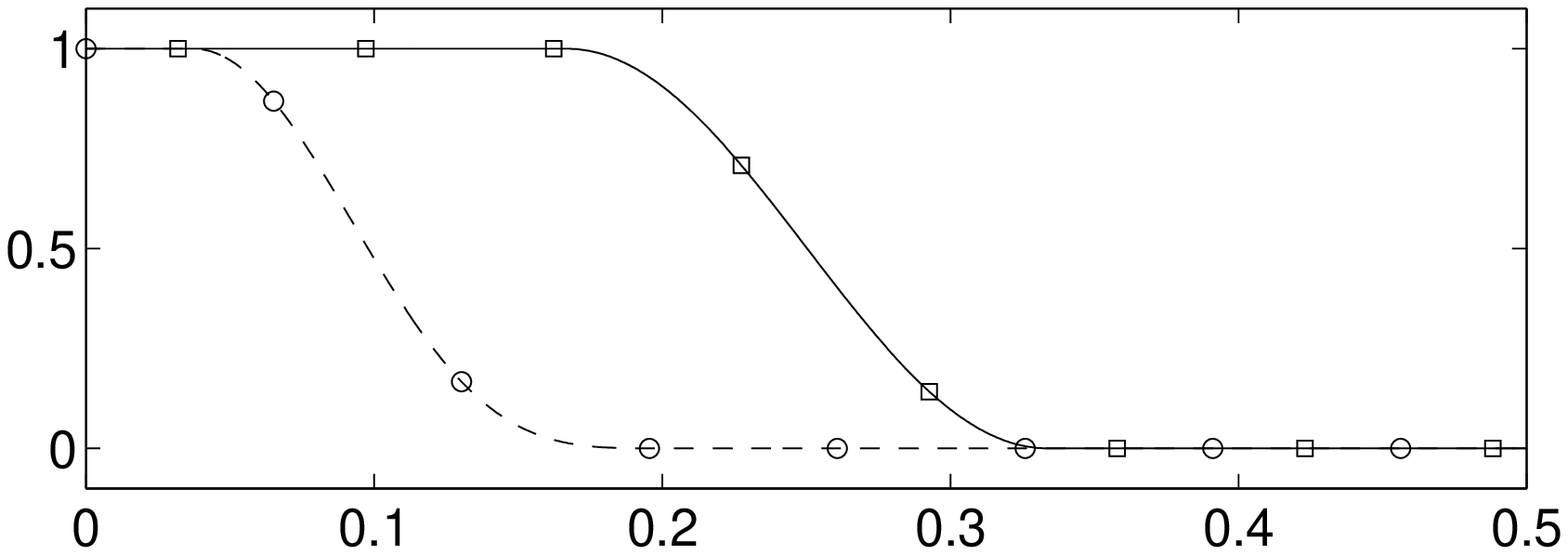}\\
\rotateleft{~~~~~~~~\rotateright{\textit{(c)}}}	& \includegraphics[width=7.5cm]{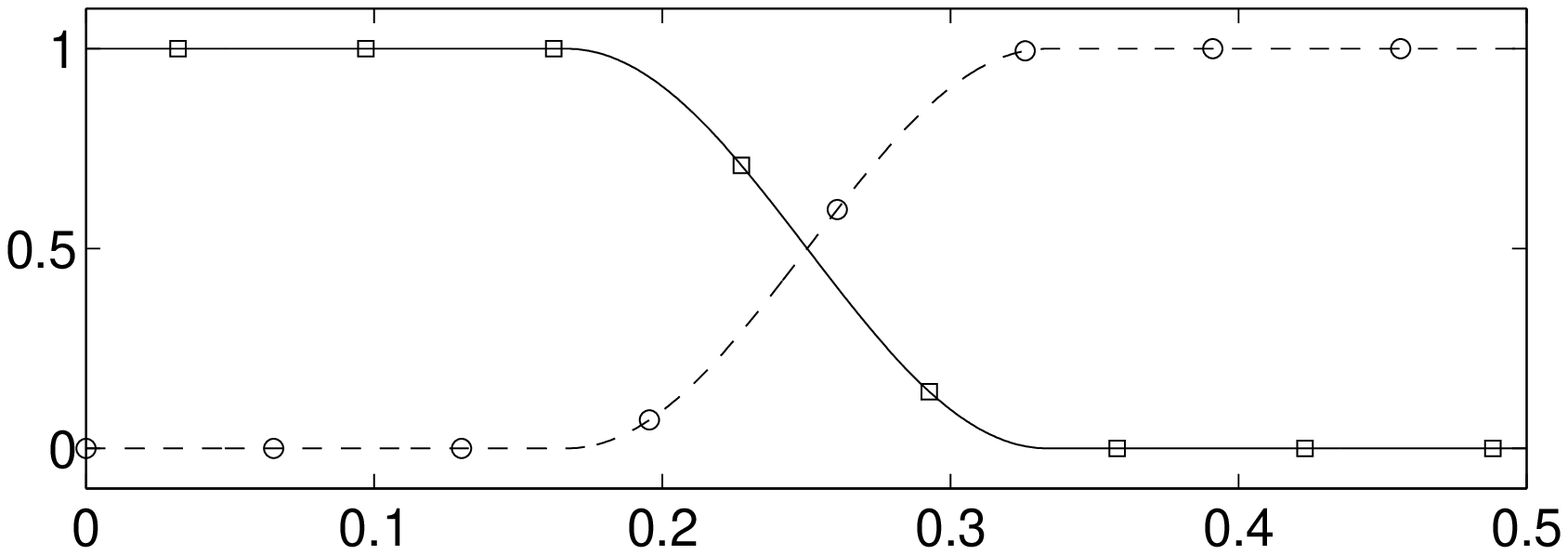}\\
 																&Reduced frequency\\
\end{tabular}
\ecc
\caption{
\textit{a}: The solid line with squares (resp. dashed line with circles) shows spectral content for ES (resp. PS). Both of them have low frequencies components. \textit{b}: The two lines show spectral contents of  correlated components (ES), with different level of correlation.
\textit{c}: Elementary decomposition for wavelet transform. The solid line with squares (resp. dashed line with circles) shows low (resp. high) frequency content.}
\label{Fig:RepartFreqWave}\label{Fig:RepartFreqBi}\label{Fig:MultiCorrel}
\end{figure}

\begin{remark} \label{Rem:Bi}
From a statistical standpoint, PS/ES can be modelized as set of uncorrelated/correlated pixels, respectively. In the Fourier plane they are respectively characterized by an extension over the whole frequency domain (PS) and an extension reduced to the low frequencies domain (ES). In particular, both of them have significant components in the low frequencies domain (see Fig.~\ref{Fig:RepartFreqBi}-a).
\end{remark}

\subsection{General bibliographical analysis}

In order to compensate for the deficiencies in the available data, additional information is (implicitly or explicitly) accounted for. Practically, most existing methods are founded on specific expected properties of observed and reconstructed sources. The proposed analysis relies on underlying decompositions of unknown image.

\medskip
\textsf{PS based methods}~--
A first part of existing methods relies on PS properties. Into this category fall original versions of CLEAN~\citep{Hogbom74,Fomalont73}, which iteratively withdraw the PS contribution to the dirty map. Early Maximum Entropy Methods (MEM) \citep{Ables74} are also founded on the properties of PS: in a regularized context, they introduce separable penalization terms (without pixel interaction) and favor high-amplitude PS. 
%

\medskip
\textsf{ES based methods}~--
Two main classes of methods have been proposed to account for the correlation of ES.

\begin{itemize}

	\item  The correlation structure is introduced by a \emph{convolution kernel}. This is the case in MEM with an Intrinsic Correlation Function (ICF)~\citep{Gull89} and Pixon methods~\citep{Dixon96,Puetter99}.

	\item The other class of method relies on \emph{pixel interactive} penalty. The early versions involve quadratic penalties \citep{Tikhonov77}. Extensions to other penalties have also been widely developed \citep{OSullivan95,Snyder92,Mugnier04}.

\end{itemize}
%

\textsf{Mixed ES+SP model}~--
The case of an \textit{explicit model mixing ES and PS} has also been addressed; however, literature in this case is poor. To our knowledge, two papers have been published: \citep{Magain98} and \citep{Pirzkal00}. They introduced the decomposition of the search map as the sum of a PS map and an ES map. From a spectral standpoint, PS\,/\,ES are respectively characterized as shown in Fig.~\ref{Fig:RepartFreqBi}-a (see also Remark~\ref{Rem:Bi}). The present paper is founded on this approach (see Sect.~\ref{Sec:BiDevelop}).

\medskip
\textsf{Multi-resolution~/~subband methods}~--
Another class of method received a large attention, namely the multi-resolution and subband approaches. 
\begin{itemize}

\item The approach proposed by \citep{Weir92,Bontekoe94} introduces structure by means of different ICF. The unknown map is the sum of several ES, with different level of correlation, \ie several low frequency components. The underlying decomposition is shown in Fig.~\ref{Fig:MultiCorrel}-b in the case of two components. 

\item We also have witnessed the development of multi-resolution extensions of CLEAN~\citep{Wakker88} as well as more subtle approaches based on wavelet decomposition and MEM~\citep{Starck94,Pantin96,Starck01}.
These methods are less specific and widely used for general deconvolution. They aim at reconstructing maps with different scales by splitting the Fourier plane into various zones. They basically rely on (recursive) decomposition in low and high frequencies as shown in Fig.~\ref{Fig:RepartFreqWave}-c. 

\end{itemize}

%
%
%
%
%

\subsection{PS plus ES: proposed developments}\label{Sec:BiDevelop}

As mentioned above, the present paper is devoted to the estimation of two distinct maps (one for ES and one for PS) from a unique set of visibilities. We then naturally resort to the work of~\citet{Magain98} and~\citet{Pirzkal00}. In both cases the PS map is written in a parametric manner founded on positions and amplitudes of peaks. Smoothness of the ES is included by means of Gaussian ICF and MEM penalty~\citep{Pirzkal00} and Tikhonov penalty~\citep{Magain98}. 
Nevertheless, they both have several limitations. On the one hand, \citep{Pirzkal00} relies on the knowledge of the position of the PS which is not available to us. On the other hand, the drawback of~\citep{Magain98} is twofold. 

\begin{enumerate}

\item It does not deconvolve with the total PSF. 

\item The optimized criterion is intricate \wrt the PS positions so, it is not always possible to find the global minimum of the criterion~\citep[ p.474]{Magain98}. 

\end{enumerate}

On the contrary, our approach achieves a complete deconvolution. Moreover, our work introduces properties so that an optimal solution is properly defined and practically attainable. In a unique coherent framework, the proposed method simultaneously accounts for intricate dirty beam, noise, the existence of point sources superimposed onto a smooth component, positivity, and the possible knowledge of a support. The estimated maps are defined as the constrained minimizer of a penalized least-squares criterion specifically adapted to this situation. So that, the method assigns a coherent value to unmeasured Fourier coefficients. The basic ideas developed here have already been partly presented within spectral analysis~\citep{Ciuciu01}, spectrometry~\citep{Djafari02a} and satellite imaging~\citep{Samson03}. 


The paper is organized as follows. In Sect.~\ref{Sec:LePb}, we define notations and state the problem in three classical forms: Fourier synthesis, spectral extrapolation\,/\,interpolation and deconvolution. All three cases concern rank-deficient linear inverse problems with additive noise. The proposed method is presented in Sect.~\ref{Sec:Regul}.  Section~\ref{Sec:Crit} introduces the regularization principles used in the subsequent sections;  Sections~\ref{Sec:MonoSharp} and~\ref{Sec:MonoExtended} respectively deal with PS and ES map; Section~\ref{Sec:Bi} is devoted to the main contribution: the reconstruction of two maps simultaneously, one consisting of PS, and the other of ES. Simulation and real data computations are presented throughout Sect.~\ref{Sec:Computations}. From a numerical optimization viewpoint, the proposed method reduces to a constrained quadratic programming problem and various options have been studied and compared. The proposed algorithm founded on augmented Lagrangian principle is presented in Sect.~\ref{Sec:Optim}. In Sect.~\ref{Sec:Conclus} we set out conclusions and perspectives.

\section{Problem statement and least squares solution} \label{Sec:LePb}

The usual model\footnote{In terms of an usual approximation, after calibration, regridding, \dots Moreover, for the sake of readability, equations are given in 1D and computation results are presented in 2D.} for the instrument writes as a weighted truncated noisy Fourier transform (discrete and regular):
\begin{equation} \label{Eq:PbDirect}
\Data = \Weight \Cvre \Fou \Objb+ \Bruit \,,
\end{equation}
where $\Objb\in\eR^N$ is the unknown map and $\Data$ and $\Bruit\in\eC^M$ are the Fourier coefficient and noise ($N$ unkown parameters for $M$ measurements).  $\Fou$ is the $N \times N$ normalized FFT matrix  and  $\Cvre$ is a $0/1$\,-\,binary  truncation (or sampling) $M \times N$ matrix ($\Cvre$ discards frequencies outside the $(u,v)$-plane coverage). $\Weight$ is a $M \times M$ diagonal matrix accounting for visibility weights. For the sake of simplicity and in accordance with real data processed in Sect.~\ref{Sec:RealDataResults}, the subsequent developments are devoted to unitary weights $\Weight=\Id_M$; they can easily be extended to include non unitary ones. Appendix~\ref{Ann:Prop} gives useful properties of these matrices. The reconstruction of $\Objb$  from $\Data$, \ie the inversion of~(\ref{Eq:PbDirect}), is a Fourier synthesis problem.

In formulation~(\ref{Eq:PbDirect}), the data $\Data$ are in the $(u,v)$-plane while the map $\Objb$ is in the image plane. Two other statements are usually given: one regarding the $(u,v)$-plane only  and the other the image plane exclusively.

\begin{enumerate}

\item In the Fourier domain, (\ref{Eq:PbDirect}) becomes a simple truncation by an invertible  change of variable $\rond{\Objb} = \Fou \Objb$:
\begin{equation} \label{Eq:PbDirectTronque}
\Data = \Cvre \rond{\Objb} + \Bruit \,.
\end{equation}
Its inversion becomes a problem of extrapolating \,/\, interpolating ``missing'' Fourier coefficients.

\item Furthermore, denoting $\bar{\Data}= \Cvre\Tr\Data$ the zero-padded data, and $\rond{\bar{\Data}} = \Fou^\dag \bar{\Data}$ the dirty map, (\ref{Eq:PbDirect}) becomes a convolution 
\begin{equation} \label{Eq:PbDirectConvol}
\rond{\bar{\Data}} = \Conv \Objb + \widetilde{\Bruit} 
\end{equation}
where $\Conv=\Fou^\dag \Cvre\Tr \Cvre\Fou $ is a (circulant) convolution matrix and $\widetilde{\Bruit}=\Cvre\Tr\Bruit$. (Superscripts ``t'' and ``$\dag$'' respectively denotes matrix transpose and conjugate-transpose).The instrument response (the dirty beam) is read in any one line of $\Conv$, up to a circular shift. The inversion becomes a deconvolution problem.

\begin{remark}
It should be noted, however, that the correlations of $\Bruit$ and $\widetilde{\Bruit}$ differ from one another, and that in this sense, the two problems are not equivalent. 
\end{remark}

\end{enumerate}

Whichever formulation is envisaged, the tackled problem is a rank-deficient linear inverse problem with additive noise. Indeed, the number of observed Fourier coefficient is far less than the number of pixels ($M\ll N$) and the operators $\Cvre \Fou$ for~(\ref{Eq:PbDirect}),  $\Cvre$ for~(\ref{Eq:PbDirectTronque}) or $\Conv$ for~(\ref{Eq:PbDirectConvol}) have $N-M$ singular values equal to~0, and $M$ singular values equal to~1. Consequently, the least-squares criterion 
\begin{equation} \label{Eq:LS}
J\LS(\Objb) = \| \Data - \Cvre \Fou \Objb \|^2 
\end{equation}
possesses an infinite number of minimizers. The dirty map is one such solution since it cancels out $J\LS$, and the other ones are obtained by adding maps with frequency components outside the $(u,v)$-plane coverage only.

\section{Regularization} \label{Sec:Regul}

So, the selection of a unique solution requires \aprio information on the searched maps to be taken into account. In order to achieve this, we resort to  regularization techniques~\citep{Idier01a,Demoment89,Tarantola87}, allowing diverse types of information to be considered, in order to exclude or avoid non desirable solutions.

\subsection{Criterion, penalization and constraints} \label{Sec:Crit}

\begin{itemize}

\item[$\bullet$] \textsf{Positivity and support}. This information is naturally encoded by hard constraints for the pixels. Let us note $\Dom$, the collection of pixels on the map, $\Supp$  the collection of pixels on a support, and $\bar{\Supp}$ its complement in $\Dom$.

	\begin{itemize}
	\item ($\mathrm{C_s}$)~: support
	\beqx
	\forall p\in\bar{\Supp}\,, ~~~ \Obj_{p}=0 \,.
	\eeqx
	The proposed method takes into account the knowledge of a support ($\Supp$ is known and $\Supp\neq\Dom$) but remains also valid if $\Supp=\Dom$.
	\item ($\mathrm{C_p}$)~: positivity
	\beqx
	\forall p\in\Dom\,, ~~~ \Obj_{p}\ge0 \,.
	\eeqx
	This information is taken to be valid in the following sections of this paper and all reconstructed maps will be positive. 
	\item ($\mathrm{C_t}$)~: template
	\beqx
	\forall p\in\Dom\,, ~~~ t_p^- \le \Obj_{p} \le t_p^+ \,.
	\eeqx
	%
	It is also possible account for a known template but it is not numerically investigated in the paper. 
	\end{itemize}

\item[$\bullet$] \textsf{Correlation structure}. Here, we are concerned with the \aprio correlation (ES) or non-correlation (PS) of the searched map. In the image plane, this information is naturally coded by penalization terms $R(\Objb)$, as a sum of potential functions $\phi$ which addresses the pixels. 

	\begin{itemize}
	\item ($\mathrm{P_c}$)~: the smooth map (ES) is favored by the introduction of interaction terms between pixels 
	\begin{equation} \label{Eq:PenalDiff}
	R_\cD(\Objb) = \sum_{p \sim q} \phi_\cD \left[\Obj_{q},\Obj_{p}\right] \,,
	\end{equation}
	%
	where $p \sim q$ symbolizes neighbor pixels.
	
	\item ($\mathrm{P_s}$)~: on the other hand, separable terms favor PS
	\begin{equation} \label{Eq:PenalSep}
	R_\sD(\Objb) = \sum \phi_\sD \left[\Obj_{p}\right] \,.
	\end{equation}
	These terms independently shrink the pixels to zero and therefore favor quasi-null maps.

	\item ($\mathrm{P_m}$)~: in the following section, we will also need to penalize the average level of the maps
	\begin{equation} \label{Eq:PenalMoy}
	R_\mD(\Objb) = \phi_\mD \left[ \sum \Obj_{p} \right] 
	\end{equation}
	so as to specifically compensate for the absence of Fourier coefficient at null-frequency.

	\item ($\mathrm{P_d}$)~: it is also possible account for a known default map $\bar{\Obj}$ through a specific penalization term such as 
	\begin{equation} \label{Eq:PenalDefautMap}
	R_\dD(\Objb) = \sum \phi_\dD \left[ \Obj_{p}, \bar{\Obj}_{p} \right] 
	\end{equation}
	but this is not numerically investigated here.

	\end{itemize}

\end{itemize}

A criterion $J$ is then introduced as a combination of some penalization terms (\ref{Eq:PenalDiff})-(\ref{Eq:PenalDefautMap}) and the data based one~(\ref{Eq:LS}) according to the objective: PS component (Sect.~\ref{Sec:MonoSharp}), ES component (Sect.~\ref{Sec:MonoExtended}) and both of them simultaneously (Sect.~\ref{Sec:Bi}).
In every case, the solution $\widehat{\Objb}$ is defined as the minimizer of $J$ under constraints $\mathrm{C_p}$ and $\mathrm{C_s}$:
\begin{equation} \label{Eq:OptimPb}
\PQSC~
\begin{cases}
~\min J(\Objb) \\ 
~~\mathrm{s.t.}
\begin{cases}
	\Obj_{p} = 0	& \mathrm{~~for~} p \in \bar{\Supp} \\
	\Obj_{p} \ge 0	& \mathrm{~~for~} p \in \Dom
\end{cases}
\end{cases}
\end{equation}
that is to say, as the solution of problem $\PQSC$. One property then becomes crucial to the construction of $J$:
\begin{itemize}
\item $(\mathrm{P_1})$~: $J$ is strictly convex and differentiable. 
\end{itemize}
Indeed, under this hypothesis,
\begin{enumerate}

\item the problem $\PQSC$ possesses a unique solution $\widehat{\Objb}$, which allows the proper definition of the estimated map;

\item the solution in question is continuous with respect to the data and to the tuning parameter values;

\item a broad class of optimization algorithms is available.

\end{enumerate}
As $J\LS$ is itself (large sense) convex and differentiable, the property $(\mathrm{P_1})$ can be assured if the potential functions are themselves convex and differentiable. Therefore, we resort to this kind of potential.

\begin{remark} \label{Rem:NonCvx}
Non-convex potentials have been introduced in image reconstruction in the 1980s~\citep{Geman84,Blake87}. As they are richer, they allow a sharper description of the searched images. For example, they can integrate binary variables, allowing contour \emph{detection} to be carried out, at the same time as image reconstruction. As a counterpart, the involved criteria can possess numerous local minima. The computational cost for optimization then increases drastically, and sometimes without guarantee against local minima. 
\end{remark}

In Sect.~\ref{Sec:Optim}, several optimization schemes have been investigated within the recommended convex framework. Various iterative algorithms solving $\PQSC$ are concerned, all of them converging to the unique solution $\widehat{\Objb}$ whatever the initialization. The only question at stake is computation time. An other property of $J$ is therefore crucial. 
\begin{itemize}
\item $(\mathrm{P_2})$~: $J$ is quadratic and circular-symmetric. 
\end{itemize}
This property allows fast optimization algorithms to be put into practice taking advantage of the FFT algorithm: fast criterion calculations, explicit intermediate solutions,\dots Since $J\LS$ is itself quadratic and circulant, $(\mathrm{P_2})$ is satisfied if the regularization terms are circulant and the potential functions $\phi$ are Quadratic~(Q) or Linear~(L). 

\begin{remark}\label{Rem:NonQuad}
Mixed convex potentials, generally quadratic about the origin and linear above a certain threshold, are used in image processing~\citep{Bouman93} and especially in astronomical imaging~\citep{Mugnier04} in order to \emph{preserve} possible edges. From the optimization strategy stand point, recent works~\citep{Idier01b,Allain04} allow to reduce the convex optimization problem to a partially quadratic one. This would make possible the development of an FFT and Lagrangian based algorithm for our PS+ES problem. We regard these forms as perspectives and we will see that forms Q and L are sufficiently rich and adapted to the envisaged contexts. 
\end{remark}

\subsection{Point Sources -- separable linear penalty} \label{Sec:MonoSharp}

This section is devoted to PS: the proposed penalization term is of type~(\ref{Eq:PenalSep}) where $\phi_\sD$ is a potential of $\eR_+$ or $\eR_+^*$ onto $\eR$, to be specified. 

Classical MEM~\citep{Nityananda82,Narayan84,Komesaroff81,Gull78,Gull84,Bhandari78,LeBesnerais99} come into play, when, for example, $\phi_\sD\left[x\right]=-\log x$, $\phi_\sD\left[x\right]=x\log x$ or $\phi_\sD\left[x\right]=-x+\bar{x}+x\log x/\bar{x}$ where $\bar{x}$ is a default map~\citep{OSullivan95,Snyder92}. They have been widely used in the domain and in image reconstruction~\citep{Djafari88c}. They have the advantage of ensuring the property ($\mathrm{P_1}$) on $\eR_+^*$, so the problem is properly regularized and $\PQSC$ possesses a unique solution. They also enjoy the advantage of ensuring (\emph{strict}) positivity, thanks to the presence of an infinite derivative at the origin $\phi_\sD'(0^+)=-\infty$~\citep{Narayan86}.

However, these functions prohibit null-pixels and this can be seen as a flaw when the searched maps are largely made up of null-pixels. On the other hand, null-pixels are favored by the introduction of a potential $\phi_\sD$ which possesses at its origin~\citep{Soussen00f}
\begin{itemize}

\item a minimum value, and

\item a strictly positive derivative.

\end{itemize}
Without loss of generality we set: $\phi_\sD(0)=0$ and $\phi_\sD'(0^+)=1$, while two possibilities allow property ($\mathrm{P_2}$) to be respected: the form L and the more general form Q. 
\beqnx
\mathrm{L}~:~~\phi_\sD(x) &=& x \\
\mathrm{Q}~:~~\phi_\sD(x) &=& \alpha x^2 + x 
\eeqnx
The penalization is then written as:
\begin{equation} \label{Eq:CritSharp}
R(\Objb) =  \lambda_\sD \sum \Obj_p + \eps_\sD \sum \Obj_p^2 \,.
\end{equation}
The strict convexity property ($\mathrm{P_1}$) imposes $\eps_\sD>0$: the~L term ensures a positive derivative at the origin, while the~Q term ensures strict convexity.

\begin{remark}
In order to favor high amplitude peaks, a least penalization function is desirable, \ie $\eps_\sD=0$. In this case, it is possible that $J$ remains unimodal or strictly convex, although we have no proof of this. This property could depend on the value of $\lambda_\sD$, on the knowledge and form of the support, on the $(u,v)$-plane coverage or on the data in each particular case. 
\end{remark}

\subsection{Extended Sources -- correlated quadratic penalty} \label{Sec:MonoExtended}

This section is devoted to ES: the penalty term of type~(\ref{Eq:PenalDiff}) introduces interactions between neighboring pixels. 

\citet{OSullivan95} proposes the use of an I-divergence: $\phi_\cD\left[\Obj,\Obj'\right] =  - \Obj + \Obj' + \Obj \log \Obj/\Obj'$or an Itakura-Saito distance: $\phi_\cD\left[\Obj,\Obj'\right] = -\log \Obj/\Obj' - 1 + \Obj/\Obj'$ in the symmetrized version. As in the case of Sect.~\ref{Sec:MonoSharp}, these allow property $(\mathrm{P_1})$ and positivity to be ensured. However, they prohibit null-pixels and do not ensure property $(\mathrm{P_2})$. 

We resort to classical terms of image processing based on finite differences between neighboring pixels. In the simplest case, first order differences yield
\beqx
\phi_\cD\left[\Obj,\Obj'\right] =  \phi_\cD\left[\Obj-\Obj'\right]
\eeqx
where $\phi_\cD$ is a potential of $\eR$ onto $\eR$ to be specified. 
In order to effectively favor smooth and correlated maps, and due to reasons of symmetry, $\phi_\cD$ is chosen to be minimal in 0 and even. In order to ensure property ($\mathrm{P_2}$), we are led to choose $\phi_\cD$ in class Q and to reject class L: $\phi_\cD(x)=x^2$ and
\beqx
R(\Objb) = \lambda_\cD \sum_{p=0}^{N} \left[\Obj_{p+1}-\Obj_p\right]^2 
\eeqx
with the hypothesis $\Obj_{0}=\Obj_{N}$ in order to ensure circularity.

We are here dealing with early regularization techniques, that appeared in the 1960s~\citep{Phillips62,Twomey63,Tikhonov63} and were developed in the mid-1970s in works by~\citet{Tikhonov77} in a continuous context and by ~\citet{Hunt77} in a discrete context. They are also related to the well-known Wiener filter.

In this form, the strict convexity condition ($\mathrm{P_1}$) is not respected. 
Indeed, $J\LS$ is not sensitive to constant maps (since null-frequency is not observed) and neither is the regularization term (since it is only a function of the difference between pixels). Several options are available for dealing with this indetermination.
\begin{enumerate}

\item Support constraint $\mathrm{C_s}$: as soon as the support constraint is valid, if at least one of the pixels is zero ($\Supp\neq\Dom$), $J$ is strictly convex on $\eR^\Supp$.

\item In the absence of support information, it is sufficient to penalize the mean of the map by a term such: 
\beqx
R_\mD(\Objb) = \left[ \sum \Obj_{p} \right]^2 \,.
\eeqx
Intuitively, it reduces the mean of the map towards 0 and is counterbalanced by the positivity constraint. 

\item It is also possible to penalize the quadratic norm of the map by a term such as that introduced in Sect.~\ref{Sec:MonoSharp}. 

\end{enumerate}
The penalization thus reads
\begin{equation} \label{Eq:CritExtended}
R(\Objb) = \lambda_\cD \sum \left[\Obj_{p+1}-\Obj_p\right]^2 	+ \eps_\mD \left[ \sum \Obj_{p} \right]^2 \,.
\end{equation}
Under this form, properties $(\mathrm{P_1})$ and $(\mathrm{P_2})$ are satisfied if ($\eps_\mD>0$\,,\,$\lambda_\cD>0$) in the case $\Supp=\Dom$ and ($\eps_\mD\ge0$\,,\,$\lambda_\cD>0$) in the case $\Supp\neq\Dom$.

\subsection{Mixed model} \label{Sec:Bi}

The present paper is devoted to maps composed of both types of component simultaneously: ES and PS. Following~\citep{Magain98} and \citep{Pirzkal00}, we introduce two maps $\ObjbES$ and $\ObjbPS$ which describe each component respectively. 
%
%
The direct model~(\ref{Eq:PbDirect}) becomes:
\begin{equation} \label{Eq:PbDirectBi}
\Data = \Cvre \Fou (\ObjbES + \ObjbPS) + \Bruit \,,
\end{equation}
and the least-squares term
\beqx
J\LSBi(\ObjbES,\ObjbPS) = \|\Data - \Cvre \Fou (\ObjbES + \ObjbPS) \|^2  \,,
\eeqx
where the subscript ``Mix'' stand for Mixed map. This form raises new indeterminates as it now concerns the estimation of $2N$ variables, still from a single set of $M$ Fourier coefficients. However, it allows the explicit introduction of characteristic information about each map through two adapted regularization terms. 

\begin{enumerate}

\item A separable term for $\ObjbPS$, identical to that in Sect.~\ref{Sec:MonoSharp}
\beqx
R_\sD(\ObjbPS) = \sum \ObjPS(p) \,,
\eeqx
minimum at 0 and with a strictly positive derivative. 

\item An interaction term between neighboring pixels of the map $\ObjbES$, identical to that in Sect.~\ref{Sec:MonoExtended}
\beqx
R_\cD(\ObjbES) = \sum \left[\ObjES(p+1)-\ObjES(p)\right]^2 \,.
\eeqx

\end{enumerate}
So as to ensure property $(\mathrm{P_1})$, the same terms as in Sect.~\ref{Sec:MonoSharp} and Sect.~\ref{Sec:MonoExtended} are added, and the regularized criterion takes the form: 
\medskip
\beqn
&& J\RegBi(\ObjbES,\ObjbPS)= J\LSBi(\ObjbES,\ObjbPS) \label{Eq:CritBi}\\
&& ~~~~+ \lambda_\sD \sum \ObjPS(p) + \eps_\sD \sum\ObjPS(p)^2 \nonumber\\
&& ~~~~+ \lambda_\cD \sum \left[\ObjES(p+1)-\ObjES(p)\right]^2 + \eps_\mD \left[ \sum\ObjES(p) \right]^2 \,,  \nonumber
\eeqn

\medskip
\noindent
where superscript ``Reg'' stands for Regularized. Re\-gu\-la\-ri\-za\-tion parameters (hyperparameters) $\lambda_\cD$ and $\lambda_\sD$ tune the smooth and spiky character of maps $\ObjbES$ and $\ObjbPS$.

In this form, properties $(\mathrm{P_1})\,-\,(\mathrm{P_2})$ are satisfied if ($\lambda_\sD\ge0, \lambda_\cD>0, \eps_\sD>0$) and  $\eps_\mD>0$ when $\Supp\neq\Dom$ or $\eps_\mD\ge0$ when $\Supp=\Dom$. The couple of maps $(\widehat{\ObjbES},\widehat{\ObjbPS})$ is properly defined as the solution of problem $\PQSC$ and the next section (Sect.~\ref{Sec:Computations}) gives the first practical results (simulated and real data processing). Section~\ref{Sec:Optim} is devoted to a fast optimization algorithm. 

\section{Computation results} \label{Sec:Computations}

\subsection{Nan\c{c}ay radioheliograph} \label{Sec:NRH}

Radio emission of the Sun at meter wavelength is known since World War II. The Nan\c{c}ay radioheliograph (NRH) is a radio-interferometer dedicated to imaging the solar corona and it monitors the radio burst in solar atmosphere at such wavelengths with high temporal rate, adequate spatial resolution and high dynamic. 

\begin{figure}[htb]
\bcc
\includegraphics[width=3.5cm]{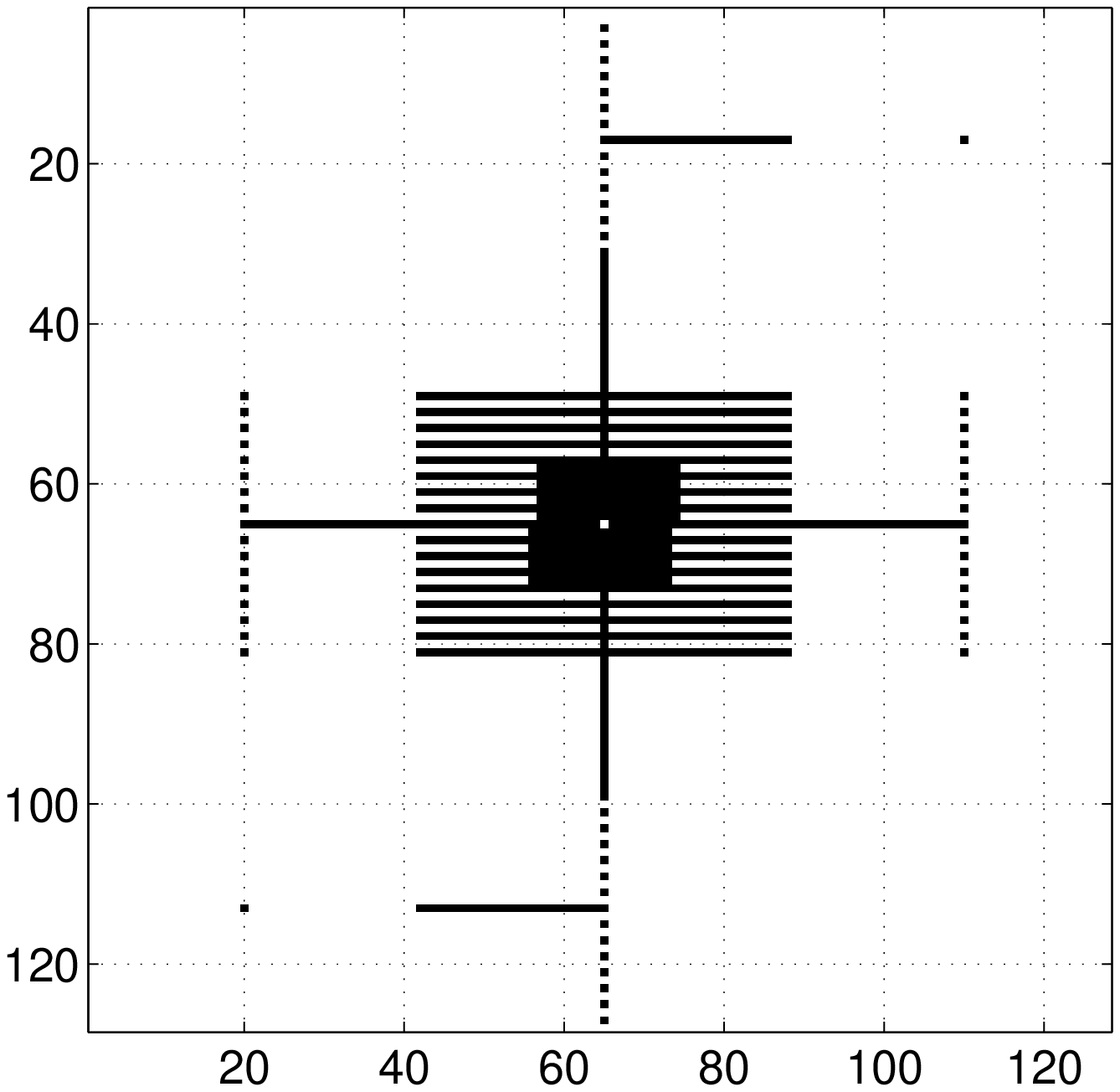} ~ \includegraphics[width=3.5cm]{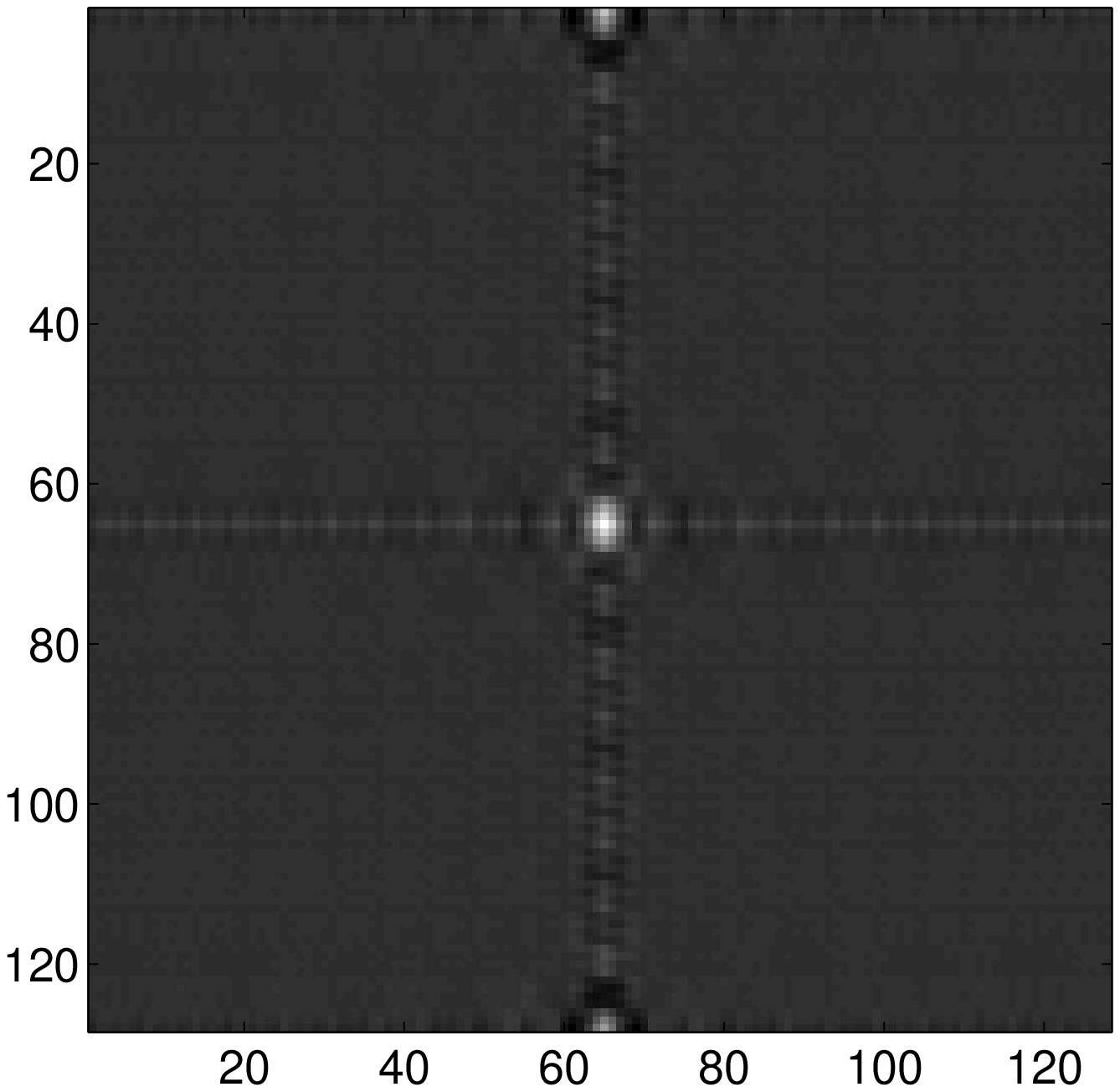}\\
\ecc
\caption[Dirty beam]{Left figure shows instantaneous $(u,v)$-plane coverage (EW array is along vertical direction and NS array is along horizontal direction). Right figure gives the dirty beam, defined as the 2D Fourier transform of the $(u,v)$-plane coverage with a unitary weight for each visibility.}
\label{Fig:CouvreEtDirtyBeam}
\end{figure}
%
%
%

At such frequencies, mainly two kinds of structures are observed in the corona: (1) larger structures (ES) and (2) smaller structures (PS). The quiet Sun (1-$i$) is the largest structure, larger than the Sun size in the visible and slowly varying on long term scale (years) \citep{Lantos96}. Medium size structures (1-$ii$) are the radio counterpart of coronal holes and magnetic loops (\emph{plateau}) \citep{Alissandrakis96}, and are also observed simultaneously in soft X rays. The time scale for such structures is days to weeks. They are clearly correlated to persistent structures observed in other wavelength (optical and X-rays) and rotate on the radio maps quasi simultaneously with their optical and X-rays counterparts. The small structures (2) with very high brightness, can often reach  several tens of Millions Kelvin \citep{Kerdraon83}; they usually have a small life time (few seconds) and are associated to energetic events in the magnetic loops in the Sun's atmosphere. Correlation with structures observed in other wavelengths is more difficult.

\begin{figure}[hbt]
\bcc
\includegraphics[width=5cm]{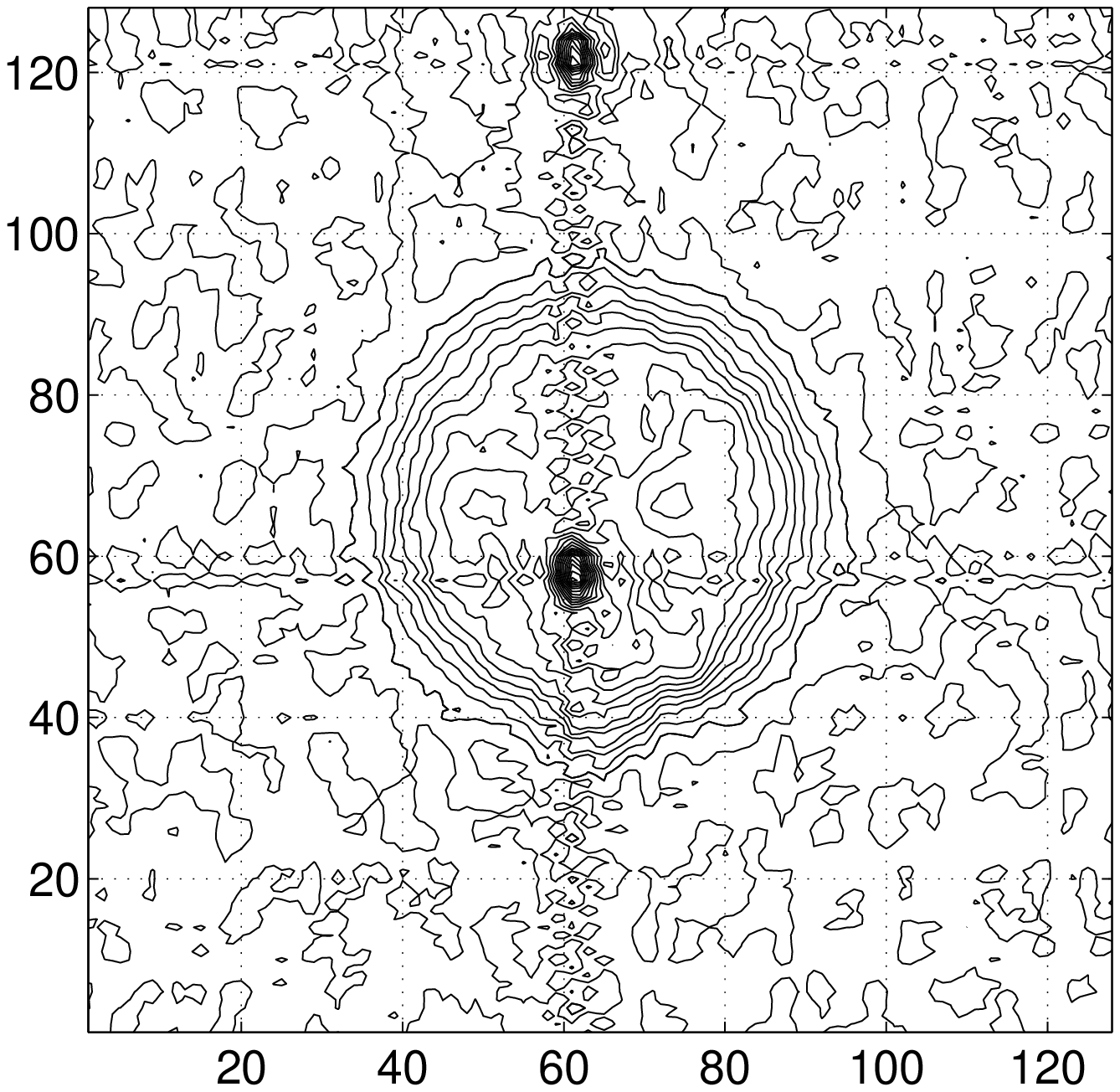}\\ 
~~\\ 
\includegraphics[width=5cm]{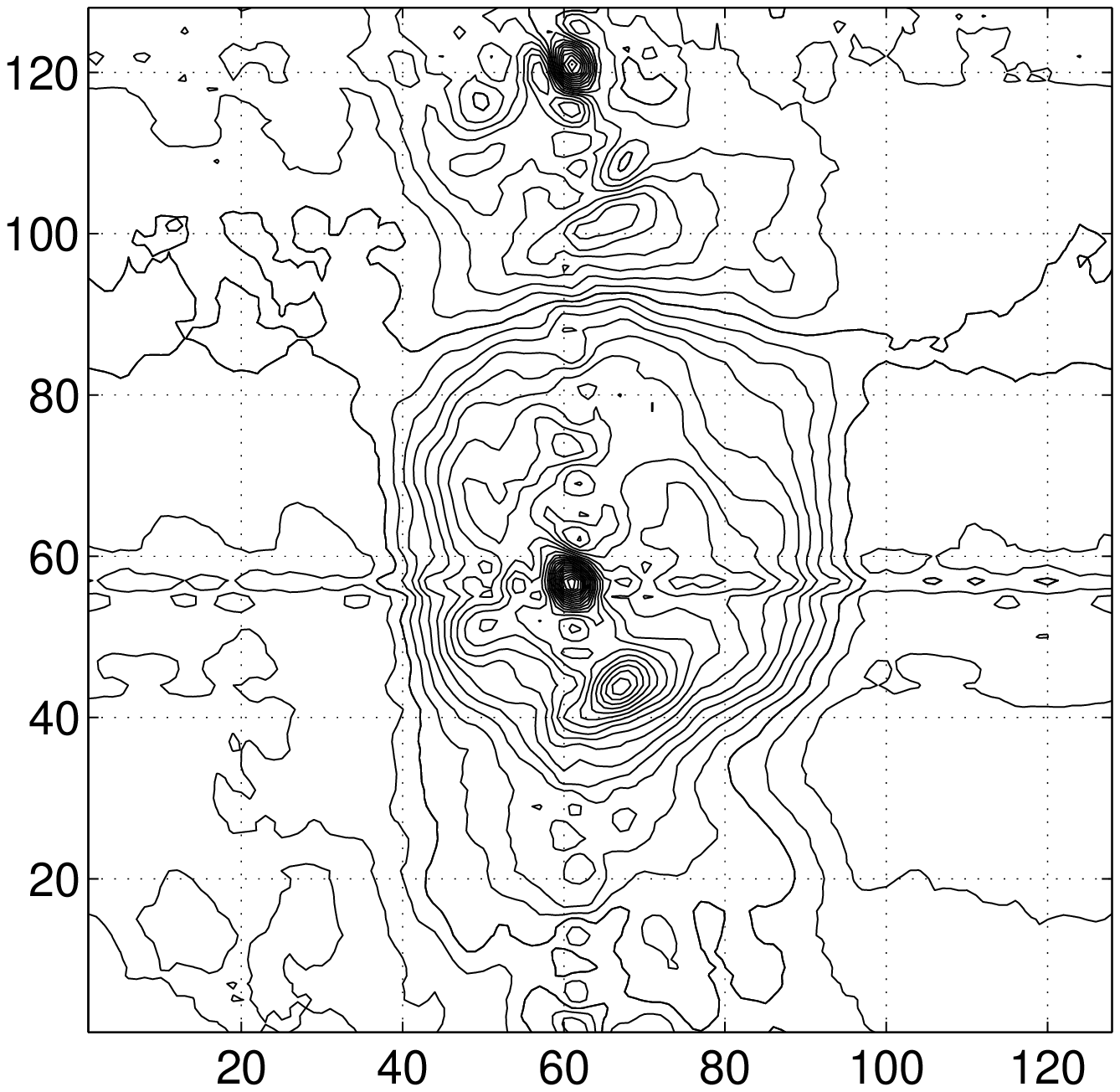}\\
\ecc
\caption[Dirty beam]{Dirty maps typically encountered with NRH: simulated data (top) and real data (bottom). Contour levels are $-10^{-2}$ to $5~10^{-2}$, step $2.5~10^{-4}$ (they are used for all the shown maps). 
}
\label{Fig:LesDirtyMaps}
\end{figure}
%
%

The NRH is composed by two arrays: one along Est-West (EW) direction with 23 antennas, the other along North-South (NS) with 19 antennas. The NRH is operating in the range 150-450\,MHz at a time sampling rate of 1/10\,s., about eight hours a day, with favorable signal to noise ratio. Since the refurbishing of the instrument \citep{Kerdraon97} cross-correlation between most of the antennas in both arrays are available. As a consequence: ($i$) 569 non redundant instantaneous visibilities are now available\footnote{Thanks to Hermitian symmetry 1138 Fourier coefficients are available. The computed $(u,v)$-plane and map are $128\times\!128$.} (with unitary weights) and moreover, ($ii$) the instantaneous coverage of the $(u,v)$-plane (shown in Fig.~\ref{Fig:CouvreEtDirtyBeam}) becomes much more uniform. Nevertheless, due to the structure of the arrays, the coverage is not uniform. The central part of the $(u,v)$-plane essentially consists of two rectangular domains: the central one is a $16\times 16$ square and the larger one is a $32\times46$ rectangle. With this configuration 2D instantaneous imaging (without Earth rotation aperture synthesis) becomes possible despite strong sidelobes in the dirty beam (see also Fig.~\ref{Fig:CouvreEtDirtyBeam}). As far as the dirty beam is concerned, the maximum value is normalized to 1 and located in the middle of the map at $(64,64)$. A secondary important lobe partly around $(1,64)$ and $(128,64)$ referred to as the aliasing lobe has amplitude $0.70$ and characterizes important aliased response. The first negative lobe is $-0.10$ around the central lobe and $-0.23$ around the aliasing lobe. Moreover, the first positive lobe is $0.14$ near the central lobe and $0.12$ near the aliasing lobe. In addition, the FWHM is $4.5$ (resp. $4$) pixels for the central (resp. aliasing) lobe. 


At processed frequency (236\,MHz), the field of view\footnote{For a declination $23\,\degr$ and null hour angle (the Sun at noon in summer), the FOV is $87\,\arcmin$ in EW and $86\,\arcmin$ in NS, and the resolution is $3.27\,\arcmin$ in EW and $2.17\,\arcmin$ in NS, since main EW arm is 1600\,m with step 50\,m and NS arm is 2640\,m with step 55\,m.} (FOV) related to the shortest baseline (55\,m in NS, 50\,m in EW) is $\sim 1\,\degr \, 20\,\arcmin $ and the size of the quiet sun is $\sim\,40\,\arcmin$, \ie $\sim$ 1/2 FOV. Since at observed  frequencies (150-450\,MHz) the FWHM of the smallest antenna primary beam (few antennas are 15\,m diameter) is much wider than the FOV, a unitary primary beam is appropriate. Moreover, the Shannon criterion is respected if the pixel number is $\sim 60$ for a FOV of $1\,\degr$.

With the given characteristics, ES/PS separation must be achieved and reconstruction errors must be as small as possible for both maps, in order to strongly constrain physical models and to monitor position, amplitude and separation of bursts. But imaging the encountered context mixing PS and ES remains difficult and standard methods such as CLEAN and MEM (even in a multiresolution approach) are usually inefficient due to the large background and the intricate mixing of real structures and sidelobes~\citep{Coulais97}. One possible outcome of the present work is to provide to the solar radio community accurate maps from NRH in order to achieve more detailed scientific studies. The following computation study (simulated and real data) is a typical case encountered with NRH and provides a first element in this sense.

\begin{figure*}[htb]
\bcc
\begin{tabular}{cc} 
\includegraphics[width=5cm]{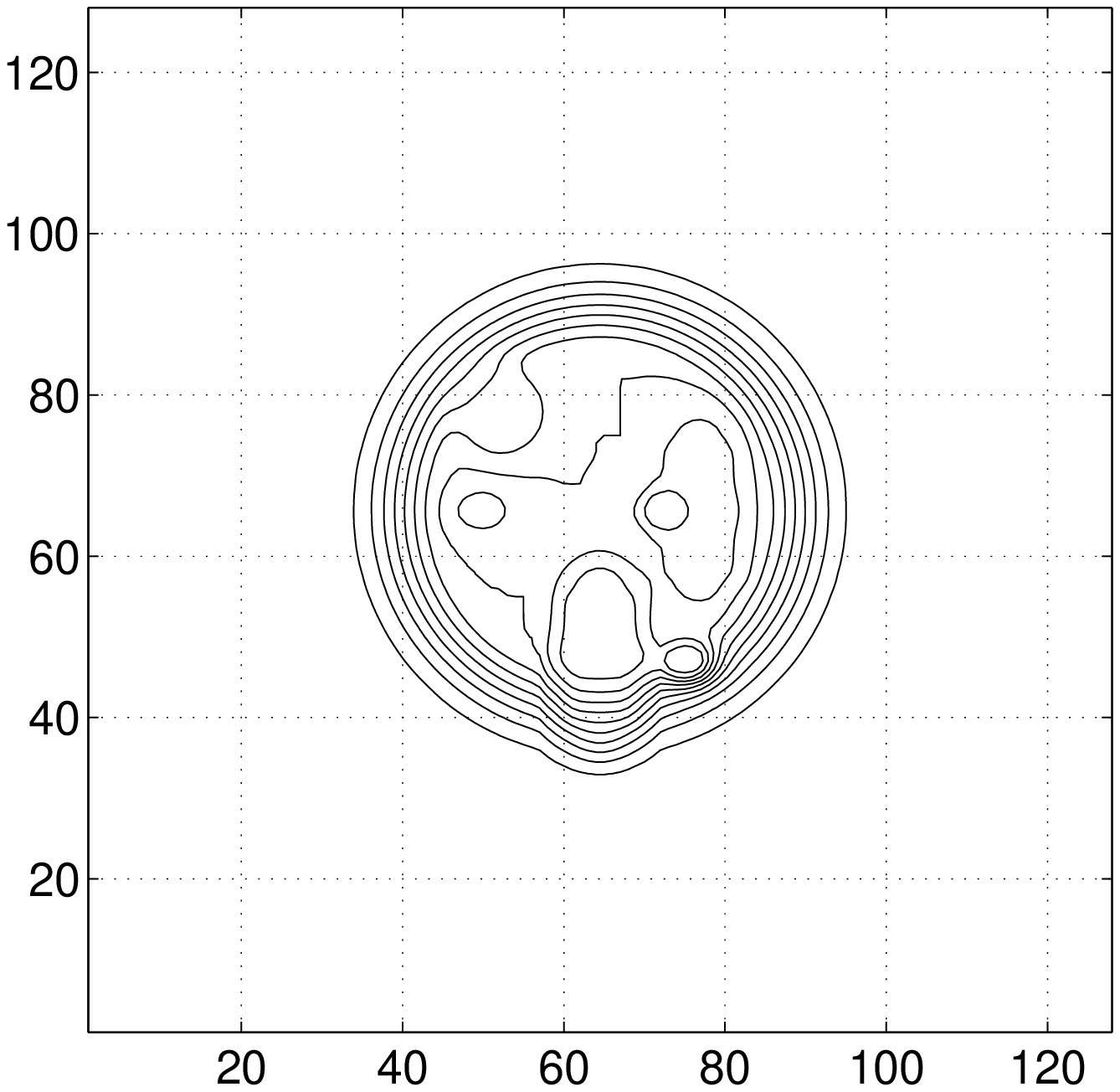} & \includegraphics[width=5cm]{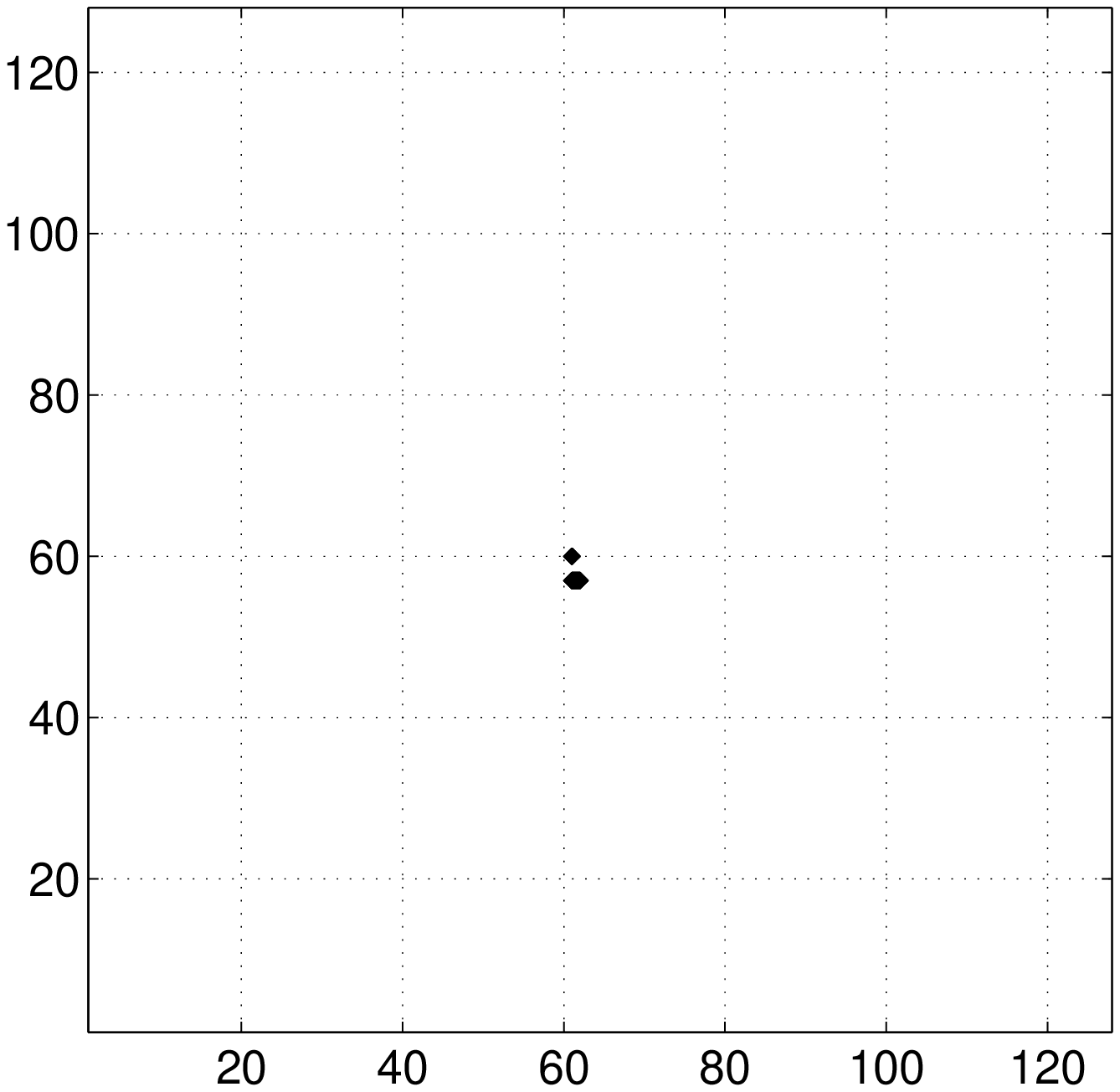}\\
 a: True object $\ObjbES^\star$& b: True object $\ObjbPS^\star$ \\
~\\
\includegraphics[width=5cm]{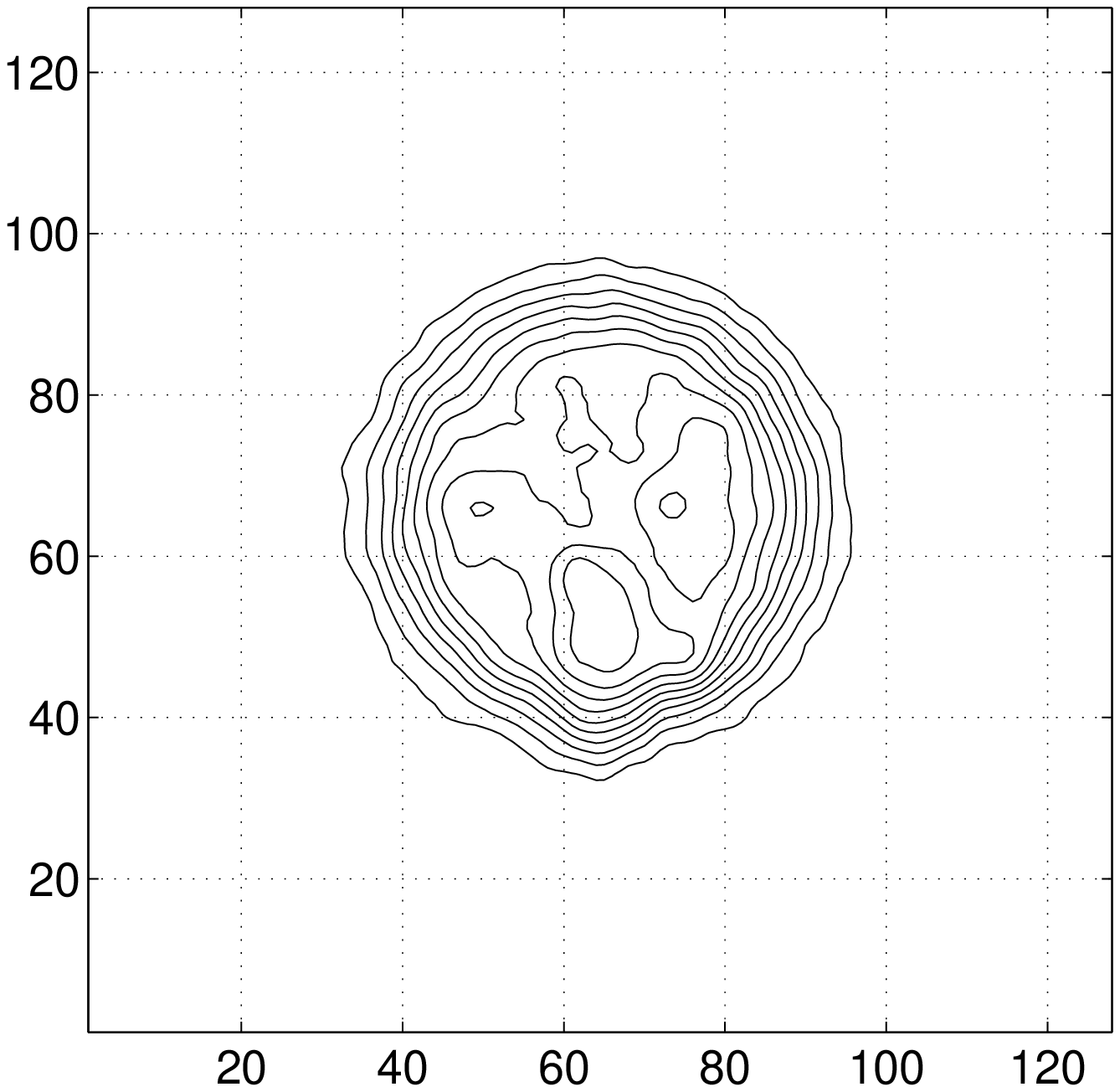} & \includegraphics[width=5cm]{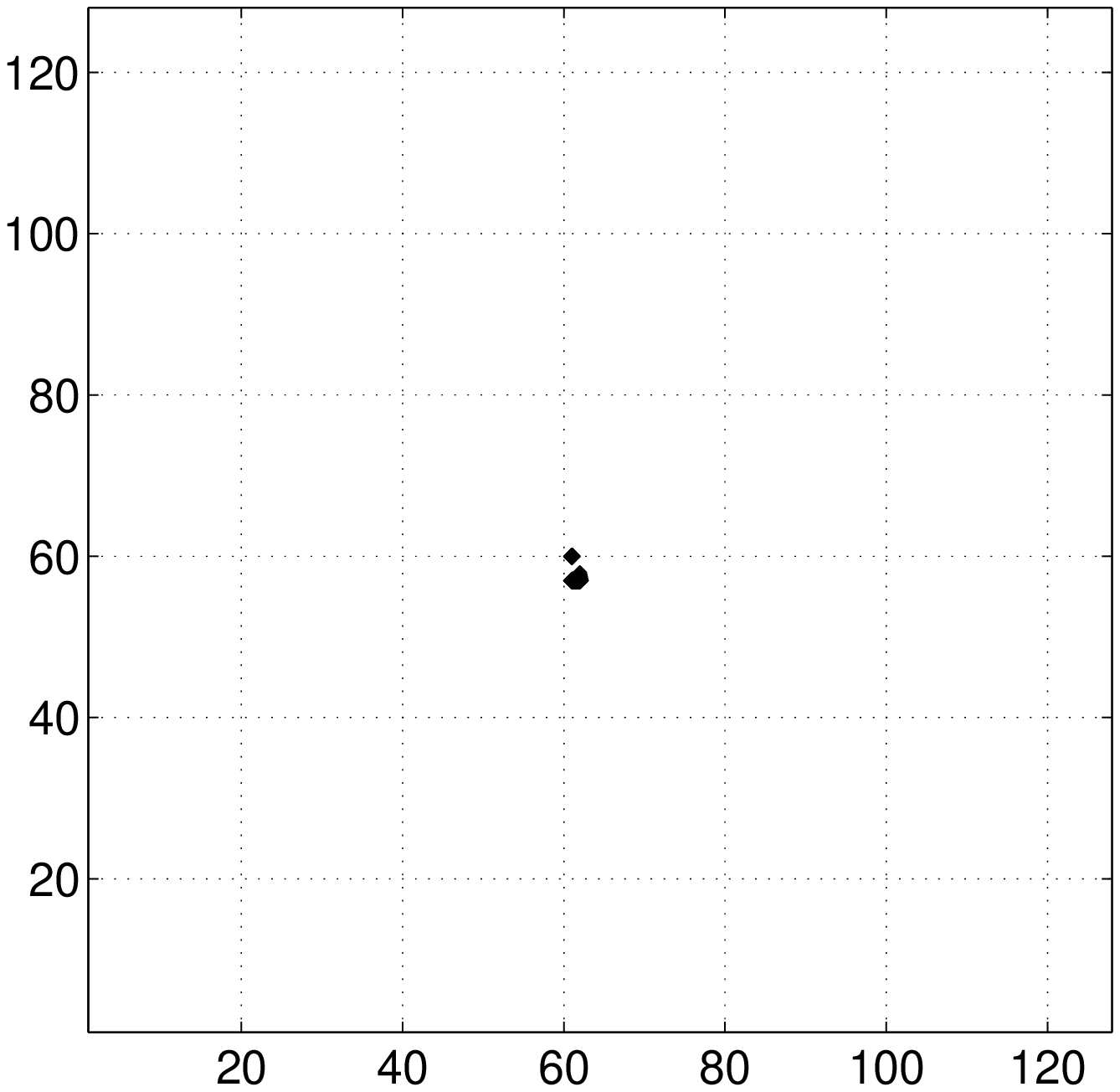}\\
c: Estimated object $\widehat{\ObjbES}$  & d: Estimated object $\widehat{\ObjbPS}$ \\
\end{tabular}
\ecc
\caption[Simulation results]{Simulation results (see Sect.~\ref{Sec:SimulDataResults}). Contour levels are the same than in Fig.~\ref{Fig:LesDirtyMaps} and~\ref{Fig:RealDataResults} for all the maps: the true ES $\ObjbES^\star$ (a) and the estimated one $\widehat{\ObjbES}$ (c) as well as for the true PS $\ObjbPS^\star$ (b) and the estimated one $\widehat{\ObjbPS}$ (d).}
\label{Fig:SimulResults}
\end{figure*}
%
\begin{figure*}[hbt]
\bcc
\begin{tabular}{cc}
\includegraphics[width=5cm]{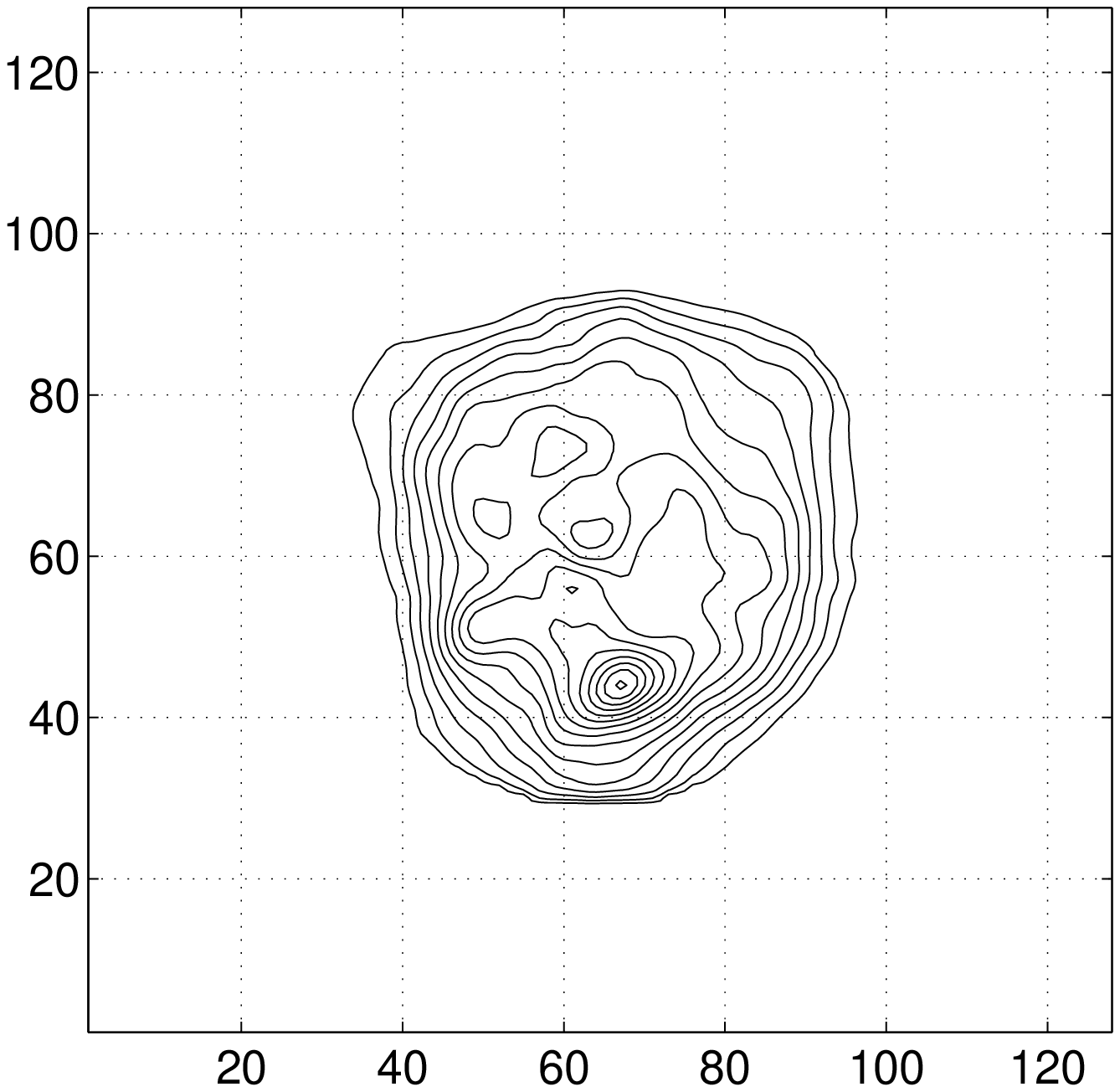} & \includegraphics[width=5cm]{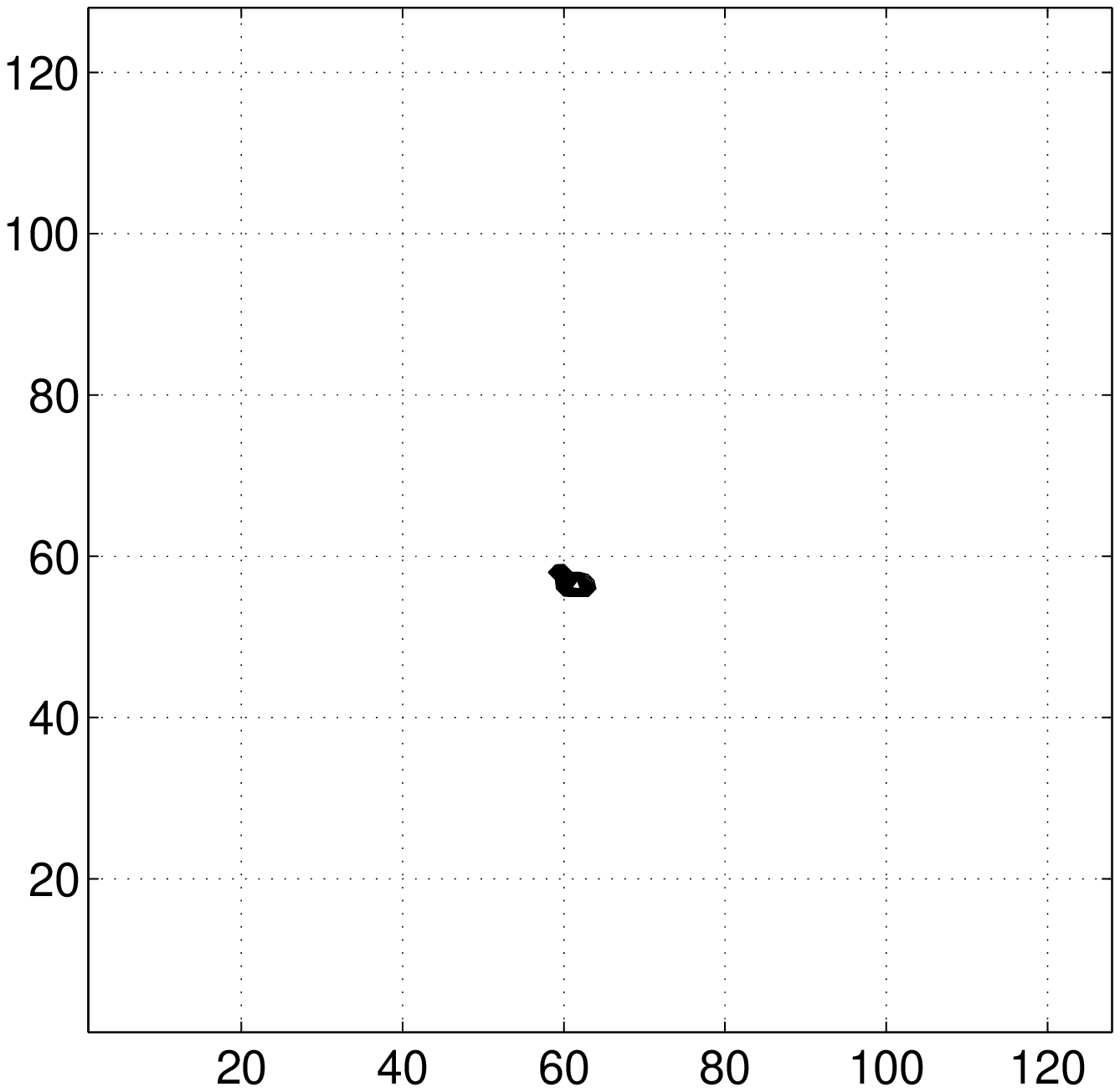}\\
a: Estimated object $\widehat{\ObjbES}$ & b: Estimated object $\widehat{\ObjbPS}$\\
\end{tabular}
\ecc
\caption[NRH data processing]{NRH data processing from typical scientific observation at 236 MHz (see Sect.~\ref{Sec:RealDataResults}). Contour levels are the same than in Fig.~\ref{Fig:LesDirtyMaps} and~\ref{Fig:SimulResults}. The two components $\widehat{\ObjbES}$ (a) and $\widehat{\ObjbPS}$ (b) are clearly separated and deconvolution of both component is clearly achieved. Both maps are positive and the prescribed supports are respected. 
}
\label{Fig:RealDataResults}
\end{figure*}

\subsection{Simulation results} \label{Sec:SimulDataResults}

\textsf{Simulated data}~

The true ES map $\ObjbES^\star$ (Fig.~\ref{Fig:SimulResults}-a), is ranging in amplitude (arbitrary units) from 0 to $5.5~10^{-3}$. The true Sun lies in a disk centered in the middle of the image, \ie  $(64,64)$ with a 64 pixels diameter. The outer part of the disk is zero and the mean of this component is $5.59~10^{-4}$. 
The true PS map $\ObjbPS^\star$ (Fig.~\ref{Fig:SimulResults}-b) consists of two peaks: the first one is located at $(60,61)$ with amplitude $5.0~10^{-2}$, and the second one overlaps pixels $(57,61)$ and $(57,62)$ with respective amplitudes  $5.0~10^{-2}$ and $4.5~10^{-2}$.
Data (in the $(u,v)$-plane) are simulated using the direct model~Eq.(\ref{Eq:PbDirectBi}), \ie FFT and truncature, and corrupted by a white, zero-mean complex Gaussian noise with variance $2~10^{-7}$. (This noise variance has been chosen in order to mimic real data).
The dirty map is shown in Fig.~\ref{Fig:LesDirtyMaps}. It is clearly dominated by the PS, and the whole map is corrupted by side lobes. Moreover, the two close peaks at location $(60,61)$ and $(57,61)-(57,62)$ are not resolved.

\medskip
\noindent
\textsf{Reconstruction parameters}~

The supports have been deduced from the dirty map. It is a disk centered at $(64,64)$ with a 70 pixels diameter for the ES map. Regarding the PS map, the support consists of one disk centered in $(58,61)$ with a 10 pixels diameter. 

In practice, two hyperparameters have to be tuned: $\lambda_\cD$ and $\lambda_\sD$ ($\eps_\sD$ is practically set to $10^{-10}$). $\lambda_\cD$ must be set in the order of magnitude of eigenvalues of the Hessian of the criterion and is set to $\lambda_\cD=2$. $\lambda_\sD$ has been empirically selected after several trials in order to visually achieve separation of PS and ES: it has been set to $\lambda_\sD=10^{-3}$.

\medskip
\noindent
\textsf{Reconstruction results}~

Fig.~\ref{Fig:SimulResults}  shows the reconstructed maps. A simple qualitative comparison with the references $\ObjbES$ and $\ObjbPS$  shows that the two components $\widehat{\ObjbES}$   and $\widehat{\ObjbPS}$  are efficiently separated and accurately reconstructed. 

The two peaks of $\widehat{\ObjbPS}$ shown in Fig.~\ref{Fig:SimulResults}-d are precisely located at $(60,61)$ and $(57,61)-(57,62)$ (overlapping). The estimated amplitudes are $0.051$, $0.048$ and $0.043$ respectively, \ie an error of less than~5\%. Moreover, the two close peaks are separated whereas they are not in the dirty map. This illustrates the resolution capability of the proposed method resulting from both data and  accounted information (positivity, support, and PS+ES hypothesis). It is also noticeable that the respective part of flux in overlapped pixels $(57,61)-(57,62)$ is correctly restored.

Fig.~\ref{Fig:SimulResults}-c gives the estimated ES map $\widehat{\ObjbES}$. Compared to the true one of Fig.~\ref{Fig:SimulResults}-a the main structures are accurately estimated. The contour lines of  Fig.~\ref{Fig:SimulResults}-c are very similar to the one of Fig.~\ref{Fig:SimulResults}-a and the relative reconstruction error is less than~2\%. Moreover, the mean of the estimated ES map $\widehat{\ObjbES}$ is $5.57~10^{-4}$ while the true mean is $5.59~10^{-4}$: the total flux is correctly estimated. The maximum value is $5.4~10^{-3}$ in $\widehat{\ObjbES}$ whereas it is $5.5~10^{-3}$ in $\ObjbES$: the dynamic is also correctly retrieved.
Nevertheless, a slight distortion located around pixel $(65,60)$ can be observed in the proposed ES map. It probably  results from an imperfect separation of the two components: a slight trace of the dirty beam remains in the estimated ES map. Moreover, the sharp edges of the true Sun are slightly smoothed due to the lack of high frequencies in the available Fourier coefficients incompletely enhanced by accounted prior information (see Remark~\ref{Rem:NonCvx} and~\ref{Rem:NonQuad}).

\subsection{Real data computations}\label{Sec:RealDataResults}

This section is devoted to real data processing based on a data set from the NRH\footnote{The eleventh of June, 2004, 13h00, at 236 MHz.}. The coverage is identical to the one of simulated data of the previous Section. 

The dirty beam is shown in Fig.~\ref{Fig:CouvreEtDirtyBeam} and the dirty map is shown in Fig.~\ref{Fig:LesDirtyMaps}. Both dirty beam and dirty map are typically encountered with NRH and are similar to the one simulated in the previous section. As expected, resolution is limited and the quality of the map is entirely contingented upon sidelobes around the brightest point sources (radio burst). Imaging such a complex context mixing PS and ES suffers from intricate mixing of real structures and sidelobes due the brightest ones. 

The same supports have been used to compute the real data and the simulated ones. It is a disk centered in the middle of the map with a 70 pixels diameter for the ES map and a disk centered in $(58,61)$ with 10 pixels diameter for the PS map. 
The same value of the parameters $\lambda_\cD=2$ and $\lambda_\sD=10^{-3}$ have been used to compute the real data and the simulated one ($\eps_\sD$ remains set to $10^{-10}$).

Estimated maps are shown in Fig.~\ref{Fig:RealDataResults}-a (ES component) and  Fig.~\ref{Fig:RealDataResults}-b (PS component). The two components $\widehat{\ObjbES}$ (Fig.~\ref{Fig:RealDataResults}-a) and $\widehat{\ObjbPS}$ (Fig.~\ref{Fig:RealDataResults}-b) are clearly separated and both deconvolution is clearly achieved. Both maps are positive and the prescribed supports are respected. Moreover, the $\widehat{\ObjbES}$ map presents a similar structure to the usual one of the Sun at meter wavelengths without strong point sources~\citep{Coulais97,Lantos96}.

\section{Numerical optimization stage} \label{Sec:Optim}

The estimated maps are defined as the unique solution of the problem $\PQSC$ given by~(\ref{Eq:OptimPb}) which involves the quadratic criterion  $J$ given by~(\ref{Eq:CritBi}). Up to an additive constant:
\beqx
J(\Objb) = \dps \frac{1}{2} ~ \Objb\Tr \,\vec{Q}\, \Objb + \vec{q}\Tr \,\Objb \,, 
\eeqx
where $\Objb=[\ObjbES;\ObjbPS]$ collects the two maps (appendix~\ref{Ann:Qq} gives $\vec{Q}$ and $\vec{q}$). Thus, $\PQSC$ is a convex quadratic program: 
\begin{equation}
\PQSC~
\begin{cases}
~\min \dps \frac{1}{2} ~ \Objb\Tr \,\vec{Q}\, \Objb + \vec{q}\Tr \,\Objb\\ \\
~\mathrm{s.t.}
\begin{cases}
	\Obj_p = 0		& \mathrm{~~for~} p \in \bar{\Supp} \\ 
	\Obj_p \ge 0	& \mathrm{~~for~} p \in \Dom
\end{cases}
\end{cases}
\end{equation}
widely investigated in the optimization literature. The main difficulty  is twofold. On the one hand, the non-separability of $J$ together with positivity constraint prevents from explicit optimization. On the other hand, the number of variables is very large. 
%
We have investigated most of the proposed methods in the excellent reference book~\citep{Nocedal00}: 

\begin{itemize}

\item Constrained gradient

\item Gradient projection

\item Barrier and interior point

\item Relaxation  (coordinate-by-coordinate)

\item Augmented Lagrangian (method of multipliers)
\end{itemize}
and have selected the latter as the faster. It is based upon successive optimizations of a Lagrangian function $\LagrFun$ founded on Lagrange multipliers $\Lagr$, slack variables $\Slack$ and quadratic penalty.  It is computationally based on FFT and threshold, so it is, in addition, very simple to implement.

\subsection{Lagrangian Function}

The equality constraint $\Obj_p=0$ ($p\in\bar{\Supp}$) is introduced by means of a usual Lagrangian term $-\Lagr_p \Obj_p$ together with a penalty term $c\Obj_p^2 /2$. 
The entire term write:
\begin{equation} \label{Eq:LagragEq}
- \sum_{p\in\bar{\Supp}} \Lagr_p \Obj_p 
	+ \frac{1}{2} \,c \sum_{p\in\bar{\Supp}} \Obj_p^2 \,.
\end{equation}

The inequality constraint $\Obj_p \ge 0$ ($p\in\Supp$) is converted into the equality one $\Slack_p -\Obj_p = 0$ using the slack variable $\Slack_p\ge 0$. Lagrange and penalty terms then write:
\begin{equation}\label{Eq:LagragIneq}
-\sum_{p\in\Supp} \Lagr_p (\Obj_p-\Slack_p) + \frac{1}{2} \,c \sum_{p\in\Supp}  (\Obj_p-\Slack_p)^2 \,.
\end{equation}

In order to simultaneously process both equality (\ref{Eq:LagragEq}) and inequality (\ref{Eq:LagragIneq}) constraints, we introduce extra slack variables $\Slack_p=0$ for $p\in\bar{\Supp}$. 
The Lagrangian then writes: 
\beqx
\LagrFun(\Objb,\Slackb,\Lagrb) = J(\Objb)
		- \Lagrb\Tr (\Objb-\Slackb) + \frac{1}{2}\,c\, (\Objb-\Slackb)\Tr (\Objb-\Slackb)
\eeqx
where $\Slackb$ and $\Lagrb$ collect slack variables $\Slack_p$ and multipliers~$\Lagr_p$. 

\subsection{Algorithm}

The algorithm then iterates three steps:
\begin{itemize}

\item[\incirc{1}] unconstrained minimization of $\LagrFun$ \wrt $\Objb$,

\item[\incirc{2}] minimization of $\LagrFun$ \wrt $\Slackb$, s.t. $\Slack_p\ge 0$,

\item[\incirc{3}] update $\Lagrb$ and $c$.

\end{itemize}
The efficiency of the proposed algorithm relies on both slack variables and property $(\mathrm{P_2})$. Roughly speaking, positivity is transfered on slack variables, so, the non separable constrained problem $\PQSC$ is split in two subproblems: a non-separable but unconstrained one computable by FFT (step \incirc{1}) and a constrained but separable one (step \incirc{2}).

\textbf{Step \incirc{1}} proceeds by fixing $\Lagrb$ and $\Slackb$ to the current value and then computes the unconstrained minimizer $\widetilde{\Objb}$ of $\LagrFun$. It is an unconstrained convex quadratic problem, so its solution is explicit:
\beqx \dps
\widetilde{\Objb} = -(\vec{Q} + c \Id_N)^{-1} 
	\left( \vec{q} + \left[ \Lagrb + c \Slackb\right]\right) \,,
\eeqx
and computable by means of FFT, thanks to circularity. 

\textbf{Step \incirc{2}} updates the slack variables $\Slack_p$ for $p\in\Supp$ (by construction, $\Slack_p=0$ for $p\in\bar{\Supp}$) as the minimizer $\widetilde{\Slack}_p$ of $\LagrFun$, subject to $\Slack_p\ge0$. 
\beqx \dps
\widetilde{\Slack}_p = 
\begin{cases}
\max{(0,c\Obj_p-\Lagr_p)}/c	& \mathrm{~~for~} p\in\Supp\\
0 										& \mathrm{~~for~} p\in\bar{\Supp}
\end{cases}
\eeqx
This step is constrained but separable: the constrained minimizer is the unconstrained one if positive and 0~if~not. 

\textbf{Step \incirc{3}} consists in updating the Lagrange multiplier~$\Lagr$: 
\beqx \dps
\widetilde{\Lagr}_p = 
\begin{cases}
\max{(0,\Lagr_p-c\Obj_p)}	& \mathrm{~~for~} p\in\Supp\\
\Lagr_p-c\Obj_p				& \mathrm{~~for~} p\in\bar{\Supp}
\end{cases}
\eeqx
This step can also include an update of $c$ (\eg $\widetilde{c}=1.1 c$). Practically, $c$ is not updated (see next subsection). 

Steps \incirc{1} to \incirc{3} are iterated until stopping condition is met,  \eg relative variation smaller that $0.1\percent$.

\begin{remark}
Constrained variables $\Obj_p=0$ for $p\in\bar{\Supp}$ can also be eliminated. This is a relevant strategy when using gradient based or relaxation methods. It does not prevent from computing $J$ and its gradient by means of FFT. On the contrary, such a strategy is not relevant here: it would break circularity and prevent from using FFT in step \incirc{1}. 
\end{remark}

\subsection{Practical case and computations time}

This section specializes the algorithm in the case of constant coefficient $c$. In this case, step \incirc{2}-\incirc{3} reduces to:
\beqx \dps
\widetilde{\Lagr_p} + c \widetilde{\Slack_p} = 
\begin{cases}
\vert \Lagr_p-c\Obj_p \vert	& \mathrm{~~for~} p\in\Supp\\
\Lagr_p-c\Obj_p					& \mathrm{~~for~} p\in\bar{\Supp}
\end{cases}
\eeqx
Moreover, $\vec{Q} + c \Id_N$ can be inverted once for all and the algorithm then requires 4 FFT per iteration. 
The algorithm has been used in the previous computations with constant coefficient $c=10^{-3}$. Convergence is achieved after about 1000 iterations and it takes  half a minute\footnote{Algorithm has been implemented with the computing environments Matlab and IDL on a PC, with a 2~GHz AMD-Athlon CPU, and 512~MB of RAM. Both codes are $\sim$ 50 lines long.}. 


\section{Conclusions} \label{Sec:Conclus}

The problem of incomplete Fourier inversion is addressed as it arises in  map reconstruction (deconvolution, spectral interpolation/extrapolation, Fourier synthesis). The proposed solution is dedicated to specific situations in which the imaged object involves two components: ($i$) an extended component together with ($ii$) a set of point sources. For this cases, new developments are given based on existing work of \citet{Magain98} and \citet{Pirzkal00}. 

The main part of the paper deals with inversion in the regularization framework. It essentially departs from usual strategies by the way it accounts for ($1$)~noise and indeterminacies, ($2$)~smoothness/sharpness prior and ($3$)~positivity and support, in a unique coherent setting. The presented development can also include known template and default map. 
Thus, a new regularized criterion is introduced and estimated maps are properly defined as its unique minimizer. The criterion is iteratively minimized by means of an efficient algorithm essentially based on Lagrange multiplier which practically requires FFT and threshold only. 
The minimizer is shown to be both practically reachable and accurate. A first evaluation of the proposed method has been carried out using simulated and real data sets. We demonstrate ability to separate the two components, high resolution capability and high quality of each map. To our knowledge, such a development is an original contribution to the field of deconvolution. 

Nevertheless, a further evaluation of the proposed method is desirable. Future work will include systematic evaluation of the capability of the proposed method as a function of $(u,v)$-plane coverage, PS amplitudes \textit{versus} ES ones, PS position (especially in a subpixelic sense) and noise level. Such an assessment concerns both simulated and real data. Moreover evaluation of the potentiality of the method on large maps (\eg VLA images), high dynamic imaging (\eg WSRT images) and imaging using millimeter interferometers (\eg IRAM PdBi and ALMA)  or optical instruments will be considered.


A part of future work in the field of SP+ES imaging, will include convex non quadratic penalization of ES (see Remark~\ref{Rem:NonQuad}).
%
%
Another part of future work will particularize the proposed method in order to produce maps of ES only and maps of PS only.

A Bayesian interpretation of the proposed method involve truncated Gauss-Markov models (ES component) and exponential white noise (PS component) and formally provides likelihood tools in order to achieve automatic tuning of the hyperparameters. This is a more delicate aspect but it will be addressed in future works.

\begin{acknowledgements}
The authors thank Anthony \textsc{Larue} and Patrick \textsc{Legros} for substantial contribution in optimization investigations. The authors are particularly grateful to Fran\c{c}ois \textsc{Viallefond}, Alain \textsc{Kerdraon}, Christophe \textsc{Marqu\'e}, Claude \textsc{Mercier} and J\'er\'emie \textsc{Lecl\`ere} for useful discussions and for providing NRH data. We are grateful to Guy \textsc{Le Besnerais} and \'Eric \textsc{Thi\'ebaut} for carefully reading the paper. Special thanks to the incredible, indescribable Gr$\ddot{\text{u}}$n.
\end{acknowledgements}

\appendix
\section{Notations} \label{Ann:Notations}


In this paper, $\Id_P$ denotes the $P\times P$ identity matrix and $\vec{M}^\dag$ (resp. $\vec{M}\Tr$) denotes the complex conjugate transpose (resp. transpose) of a given matrix  $\vec{M}$. 
%

Let us note $\Doux$ the (circulant) first order difference matrix and $\Lambdab = \Fou \Doux \Fou^\dag$ the diagonalized matrix. 
Let us also note $\Unbb$ the ones column vector with $N$ components and $\rond{\Unbb}= \Fou\Unbb$ its FFT (non-null at null frequency only). 

We introduce now two matrices $\DeltaE$ and $\DeltaP$ useful to compute ES and PS respectively: 
%
%
\beqx
\begin{cases}
\DeltaE = \DeltaC + \lambda_\cD \, \Lambdab  + \eps_\mD \, \rond{\Unbb}^\dag \rond{\Unbb}	\label{Eq:DeltaE}\\
\DeltaP = \DeltaC +  \eps_\sD \, \Id_N \label{Eq:DeltaP}
\end{cases}
\eeqx
where $\DeltaC=\Cvre\Tr \Cvre$. The three sub-matrices $\DeltaE$, $\DeltaP$ and $\DeltaC$ are diagonal matrices.

\section{Conventions and properties} \label{Ann:Prop}


This Appendix gives several properties of $\Fou$ and $\Cvre$ introduced in Sect.~\ref{Sec:LePb}.
\begin{itemize}

\item $\Fou^\dag \Fou = \Fou\Fou^\dag =\Id_N$~: orthonormality of the normalized FFT. 

\item $\Cvre$ is a truncation operator, $N\times M$, (eliminates coefficient outside the coverage). 

\item $\Cvre\Tr$ is a zero-padding operator, $M\times N$, (adds null coefficient outside the coverage). 

\item $\DeltaC = \Cvre\Tr \Cvre$ is a projection matrix, $N\times N$, (nullifies coefficients outside the coverage). 

\item $\Cvre \Cvre\Tr = \Id_M$.

\end{itemize}

\section{Gradient and Hessian calculi} \label{Ann:Qq}

\renewcommand{\arraystretch}{1.5}
\begin{table*}[bt]
\begin{equation}
\begin{array}{cccccc} \hline
\dps \rho(\Objb)													~~~&
\dps \rond{\rho} (\rond{\Objb})								~~~&
\dps \partial \rho / \partial \Objb							~~~&
\dps \partial \rond{\rho} / \partial \rond{\Objb}		~~~&
\dps \partial^2 \rho / \partial \Objb^2					~~~&
\dps \partial^2 \rond{\rho} / \partial \rond{\Objb}^2	~~~\\ \hline
\|\Data - \Cvre \Fou \Objb \|^2								~~~&
\|\Data - \Cvre \rond{\Objb} \|^2							~~~&
- 2\, \Fou^\dag \Cvre\Tr (\Data - \Cvre\Fou \Objb )	~~~&
- 2\, \Cvre\Tr (\Data - \Cvre \rond{\Objb} )				~~~&
2\, \Fou^\dag \Cvre\Tr  \Cvre\Fou							~~~&
2\, \Cvre\Tr  \Cvre												~~~\\
\Objb\Tr \Doux\Tr \Doux \Objb									~~~&
\rond{\Objb}^\dag \Lambdab^\dag\Lambdab \rond{\Objb}	~~~&
2\, \Doux\Tr \Doux \Objb										~~~&
2\, \Lambdab^\dag\Lambdab \rond{\Objb}						~~~&
2\, \Doux\Tr \Doux												~~~&
2\, \Lambdab^\dag\Lambdab										~~~\\
\Objb\Tr\Objb														~~~&
\rond{\Objb}^\dag  \rond{\Objb}								~~~&
2\, \Objb															~~~&
2\, \rond{\Objb}													~~~&
2\, \Id_N															~~~&
2\, \Id_N															~~~\\
(\Unbb\Tr \Objb)^2												~~~&
\rond{\Objb}(0)^2													~~~&
2\, \Unbb \Unbb\Tr \Objb										~~~&
2\, \rond{\Unbb} \rond{\Unbb}^\dag \rond{\Objb}			~~~&
2\, \Unbb \Unbb\Tr												~~~&
2\, \rond{\Unbb}^\dag \rond{\Unbb}							~~~\\
\Unbb\Tr \Objb														~~~&
\rond{\Objb}(0)													~~~&
\Unbb																	~~~&
\rond{\Unbb}														~~~&
\vec{0}																~~~&
\vec{0}																~~~\\ \hline
\end{array}
\end{equation}
\caption{Functions, gradients and Hessian of encountered criteria (given as a function of map and their FFT).}
\end{table*}

This Appendix is devoted to the vector $\vec{q}$ and the matrix $\vec{Q}$ involved in the minimized criterion. 

$\vec{q}$ is a $2N$ components column vector based on the gradient of criterion $J$, at $\Objb=0$. The first part is the dirty map and the second one is the dirty map minus a constant map equal to $\lambda_\sD/2$. In the Fourier domain, $\vec{q}$ reads:
 
\beqx
\rond{\vec{q}} = \left. \frac{\partial \rond{J}}{\partial \rond{\Objb}} \right|_{\Objb=0} =
\left[\begin{array}{l}
\dps\frac{\partial \rond{J}}{\partial \rond{\ObjbES}} \\ \\
\dps\frac{\partial \rond{J}}{\partial \rond{\ObjbPS}} 
\end{array}\right]_{\Objb=0}
=
-2 \left[\begin{array}{l}
\dps \bar{\Data}\\ 
\dps \bar{\Data} - \lambda_\sD \rond{\Unbb} /2
\end{array}\right] \,.
\eeqx


$\vec{Q}$ is a $2N\times 2N$ matrix based on the Hessian of~$J$. The two anti-diagonal elements are the Hessian of the LS term and rely on the dirty beam only. The diagonal elements are the Hessian of $J$ \wrt each map $\ObjbPS$ and $\ObjbES$. In the Fourier domain, $\vec{Q}$ reads:
\beqx
\rond{\vec{Q}} = \frac{\partial^2 \rond{J}}{\partial \rond{\Objb}^2} = 
\left[\begin{array}{cc}
\dps\frac{\partial^2 \rond{J}}{\partial\rond{\ObjbES}^2}	& \dps\frac{\partial^2 \rond{J}}{\partial\rond{\ObjbES} ~ \partial\rond{\ObjbPS}} \\
&\\
\dps\frac{\partial^2 \rond{J}}{\partial\rond{\ObjbPS} ~ \partial\rond{\ObjbES}} & \dps \frac{\partial^2 \rond{J}}{\partial\rond{\ObjbPS}^2} 
\end{array}\right]
=
\left[\begin{array}{cc}
\DeltaE~&~\DeltaC \\
\DeltaC~&~\DeltaP
\end{array}\right] \,.
\eeqx
%

\section{Object updates }

The present subsection gives details about the step \incirc{1} of the proposed algorithm (Sect.~\ref{Sec:Optim})~: the unconstrained minimization of $\LagrFun$ \wrt $\Objb$, \ie the update of $\ObjbES$ and $\ObjbPS$. Let us introduce the two vectors 
%
\beqx
\begin{cases}
\vec{z}_\eD=\bar{\Data} + (\rond{\Lagrb}_\eD + c\,\rond{\Slackb}_\eD)/2 \\
\vec{z}_\pD=\bar{\Data} + (\rond{\Lagrb}_\pD + c\,\rond{\Slackb}_\pD)/2  - \lambda_\sD \, \rond{\Unbb}
\end{cases}
\eeqx
based on observed data $\bar{\Data}$ and FFT of slack variables and Lagrange multipliers $\rond{\Slackb}=\Fou\Slackb$ and $\rond{\Lagrb} = \Fou \Lagrb$ (for each map ES and PS). Let us also introduce two diagonal matrices
\beqx
\begin{cases}
\MatE = \DeltaE + c \, \Id_N /2\\
\MatP = \DeltaP + c \, \Id_N /2 
\end{cases}
\eeqx
The update reads~:
\beqx
\begin{cases}
\rond{\ObjbES} = \left( \MatE \, \MatP - \DeltaC^2 \right)^{-1} \left( \MatP ~\vec{z}_\eD - \DeltaC ~\vec{z}_\pD \right)\\
\rond{\ObjbPS} = \left( \MatE \, \MatP - \DeltaC^2 \right)^{-1} \left( \MatE ~\vec{z}_\pD - \DeltaC ~\vec{z}_\eD \right)
\end{cases}
\eeqx
%
easily implemented since $\MatE \, \MatP - \DeltaC^2$ is diagonal.

\def\sca{\textsc}

\bibliographystyle{aa}              

\end{document}